\PassOptionsToPackage{table,xcdraw,dvipsnames}{xcolor}
\documentclass[acmtog, nonacm]{acmart}
\acmSubmissionID{172}

\usepackage[table,xcdraw,dvipsnames]{xcolor}

\usepackage{amsmath, mathtools, bm, cancel, calc, accents}
\usepackage{booktabs, tabularx, makecell, wrapfig}
\usepackage{graphicx}
\usepackage{rotating}
\usepackage{algpseudocode}
\usepackage{units} 

\usepackage{subcaption}

\usepackage{tikz}
\usetikzlibrary{
  backgrounds,
  decorations.pathreplacing,
  arrows.meta,
  tikzmark,
  calc,
  shapes.geometric,
  fit
}
\usepackage{tikzscale}
\usepackage{pgfplots}
\usepgfplotslibrary{groupplots}
\pgfplotsset{compat=1.18}

\usepackage{cleveref}

\newcommand\Revision[1]{{\color{black}{#1}}}

\tikzset{every node/.style={font=\fontfamily{LinuxBiolinumT-TLF}\selectfont}}
\usepackage{bm}
\usetikzlibrary{spy}

\newcommand{\reffig}[1] {Figure~\ref{#1}}

\newcommand{\zhecheng}[1]{}

\newlength{\subcolumnwidth}
\newenvironment{subcolumns}[1][0.45\columnwidth]
 {\valign\bgroup\hsize=#1\setlength{\subcolumnwidth}{\hsize}\vfil##\vfil\cr}
 {\crcr\egroup}
\newcommand{\nextsubcolumn}[1][]{%
  \cr\noalign{\hfill}
  \if\relax\detokenize{#1}\relax\else\hsize=#1\setlength{\subcolumnwidth}{\hsize}\fi
}

\usepackage[ruled]{algorithm2e} 

\SetAlFnt{\small}
\SetAlCapFnt{\small}
\SetAlCapNameFnt{\small}
\SetAlCapHSkip{0pt}

\acmJournal{TOG}

\copyrightyear{2025}
\acmYear{2025}
\setcopyright{acmlicensed}\acmConference[SIGGRAPH Conference Papers '25]{Special Interest Group on Computer Graphics and Interactive Techniques Conference Conference Papers }{August 10--14, 2025}{Vancouver, BC, Canada}
\acmBooktitle{Special Interest Group on Computer Graphics and Interactive Techniques Conference Conference Papers (SIGGRAPH Conference Papers '25), August 10--14, 2025, Vancouver, BC, Canada}
\acmDOI{10.1145/3721238.3730597}
\acmISBN{979-8-4007-1540-2/2025/08}

\makeatletter
\newcommand{\showfontinfo}{
    Font: \f@family \f@series \f@shape, Size: \f@size
}
\makeatother
\begin{document}
\title{Lifting the Winding Number: \\ Precise Discontinuities in Neural Fields for Physics Simulation}

\author{Yue Chang}
\orcid{0000-0002-2587-827X}
\affiliation{%
  \institution{University of Toronto}
  \country{Canada}}
\email{changyue.chang@mail.utoronto.ca}

\author{Mengfei Liu}
\orcid{0000-0003-4989-6971}
\affiliation{%
  \institution{ University of Toronto}
  \country{Canada}}
\email{mengfei.liu@mail.utoronto.ca}

\author{Zhecheng Wang}
\orcid{0000-0003-4989-6971}
\affiliation{%
  \institution{ University of Toronto}
  \country{Canada}}
\email{zhecheng@cs.toronto.edu}

\author{Peter Yichen Chen}
\orcid{0000-0003-1919-5437}
\affiliation{%
  \institution{MIT CSAIL}
  \country{USA}}
\email{pyc@csail.mit.edu}

\author{Eitan Grinspun}
\orcid{0000-0003-4460-7747}
\affiliation{%
  \institution{ University of Toronto}
  \country{Canada}}
\email{eitan@cs.toronto.edu}

\begin{abstract}
Cutting  thin-walled deformable structures is common in daily life, but poses significant challenges for simulation due to the introduced spatial discontinuities. Traditional methods rely on mesh-based domain representations, which require frequent remeshing and refinement to accurately capture evolving discontinuities. These challenges are further compounded in reduced-space simulations, where the basis functions are inherently geometry- and mesh-dependent, making it difficult or even impossible for the basis to represent the diverse family of discontinuities introduced by cuts.

Recent advances in representing basis functions with neural fields offer a promising alternative, leveraging their discretization-agnostic nature to represent deformations across varying geometries. However, the inherent continuity of neural fields is an obstruction to generalization, particularly if discontinuities are encoded in neural network weights.

We present \emph{Wind Lifter}, a novel neural representation designed to accurately model complex cuts in thin-walled deformable structures. Our approach constructs neural fields that reproduce discontinuities precisely at specified locations, without ``baking in'' the position of the cut line. To achieve this, we augment the input coordinates of the neural field with the generalized winding number of any given cut line, effectively lifting the input from two to three dimensions. Lifting allows the network to focus on the easier problem of learning a 3D everywhere-\emph{continuous} volumetric field, while a corresponding restriction operator enables the final output field to precisely resolve \emph{strict} discontinuities. Crucially, our approach does not embed the discontinuity in the neural network's weights, opening avenues to generalization of cut placement.

Our method achieves real-time simulation speeds and supports dynamic updates to cut line geometry during the simulation. Moreover, the explicit representation of discontinuities makes our neural field intuitive to control and edit, offering a significant advantage over traditional neural fields, where discontinuities are embedded within the network’s weights, and enabling new applications that rely on general cut placement.
    
\end{abstract}

%
%


%
%


\begin{teaserfigure}
    \centering
    \includegraphics[width=\linewidth]{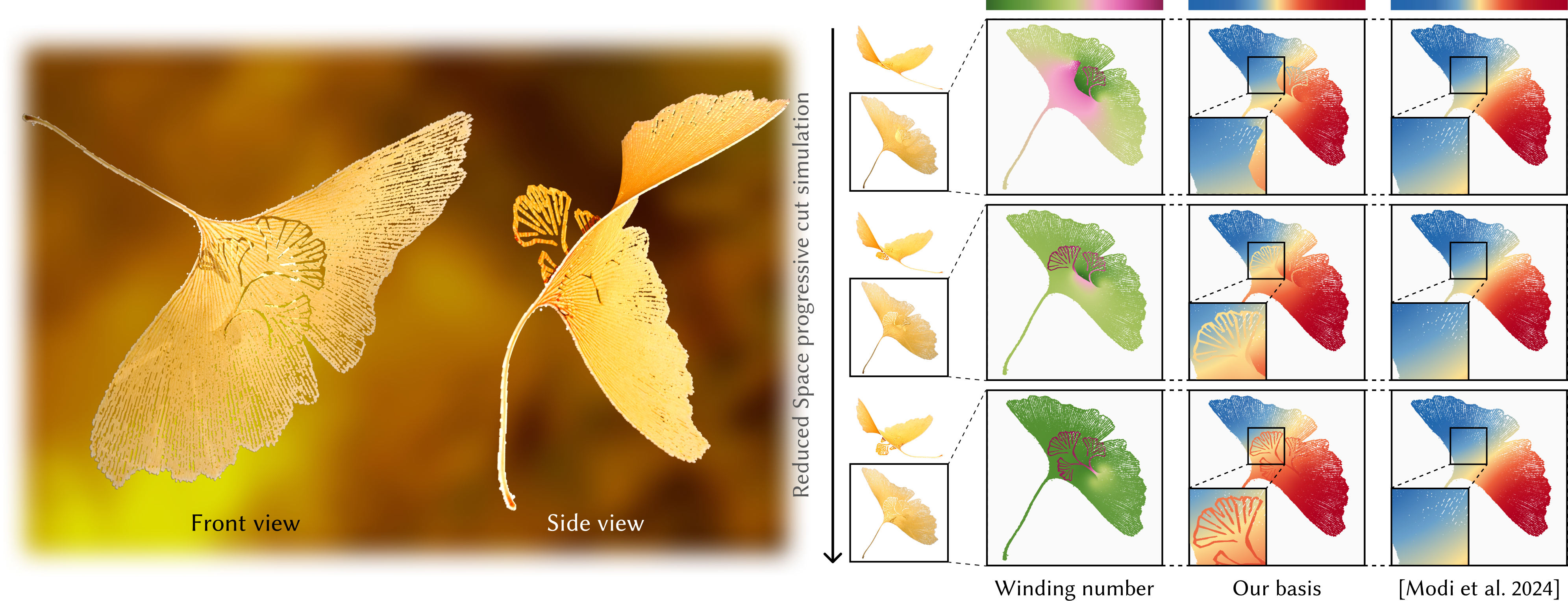}
    \caption{Our method calculates a basis for reduced-space simulations that can represent shapes with a family of discontinuities. This basis is represented by neural fields. Unlike neural fields implemented by vanilla MLPs \cite{Modi:2024:Simplicits}, which have built-in continuity, we represent discontinuities by construction using the winding number field. The basis can then be used for reduced-space simulation for progressive cutting.}
    \label{fig:teaser}
\end{teaserfigure}

\maketitle

\section{Introduction}

\label{sec:introduction}
From laser-cut panels to handcrafting shapes with scissors, cuts pervade daily life. Simulating cuts in thin-walled structures such as leaves, paper, or fabric requires a kinematic representation that precisely resolves time-evolving spatial discontinuities. This is a requirement that is not met by current reduced order modeling (ROM) approaches \cite{barbivc2005real}.

\begin{figure*}
\includegraphics[width = \linewidth]{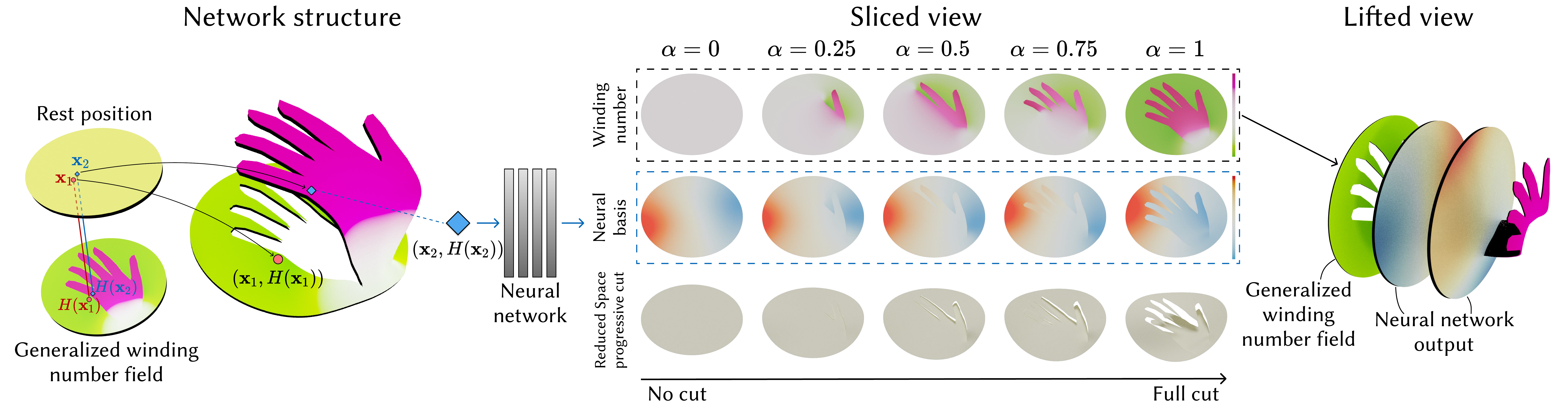}
\vspace{-20pt}
\caption{We augment the input to our neural field with a generalized winding number, lifting the domain from 2D into 3D. As the cut progresses, the generalized winding number changes accordingly throughout the domain, capturing the family of discontinuities. The neural network takes the lifted input and outputs the basis function. This allows the neural network to learn a continuous function. The discontinuity is captured by the restriction of a continuous function over a 3D domain.}
\label{fig:hand}
\end{figure*}

ROM is an acceleration technique for physical simulations that is applicable for scenarios where the \emph{deformation} complexity is low relative to the \emph{geometric} complexity. In ROM, an offline precomputation identifies a truncated set of kinematic modes, which are leveraged in fast ``reduced'' simulation. By computing the dynamics of only a small number of modes, in lieu of a large number of geometric (e.g., mesh) degrees of freedom, ROMs offer speedups by orders of magnitude~\cite{benner2015survey}. 

However, ROM's benefits come at the cost of \emph{precision} and \emph{generalization}. ROMs are currently unable to treat cutting of thin-walled materials, particularly when cut placement may vary, cutting is progressive, or the cut boundary is complex. And yet, since detailed cuts introduce considerable geometric complexity (see Fig.~\ref{fig:teaser}), much could be gained if ROM could separate kinematics from geometry.

To effectively model cuts, ROM simulations would need a novel kinematic basis that spans the deformations induced by cutting and loading the deformable object. ROMs struggle with this, because cutting induces \emph{strict} discontinuities in otherwise continuous displacement fields. 

Current ROMs either lack representation or ``bake in'' the placement of discontinuities, therefore, they do not generalize over cut placement. For instance,  mesh-based ROM mode precomputation ties the basis to the underlying discretization  \cite{sifakis2012fem,Fulton:LSD:2018,shen2021high}. Since progressing or altering a cut changes mesh connectivity, these ROMs cannot reuse the learned basis and must again precompute offline.


Recent work has leveraged continuous neural fields as kinematic basis representations \cite{pan2022neural,chen2023crom,puri2024snf,tao2024neural}. The advantage of these methods is that their precomputation is agnostic to the specific discretization of the domain geometry. However, current neural field techniques have shown limited capacity for representing high-frequency or discontinuous data~\cite{Belhe:2023:DiscontinuityAwareNeuralFields}.

Indeed, efforts to represent discontinuous displacements in neural reduced simulations are nascent. Recently, \citet{chang:2023:licrom} trained on simulation snapshots of a small, manually-selected set of partial cuts, producing a displacement basis that encodes the discontinuities of partial cuts in the neural network's weights. This limits their online reduced simulations to only the cut position seen during training, and limits the precision of the cut boundary due to the neural network's limited capacity to precisely represent a discontinuity along a curve.

\paragraph{Contributions}
We propose a novel neural field construction and ROM approach specifically designed to \emph{precisely} capture \emph{strict} displacement discontinuities for a family of progressively-cut shapes over 2D domains. Our method enables reduced simulations of progressive cutting, for cut placements not seen during training. It embraces both training on existing deformation \emph{data}, or training \emph{data-free} by identifying natural vibration modes. The resulting reduced simulation maintains an explicit representation of cut positions as polylines that may be freely modified as the simulation proceeds, without remeshing or similar data structure updates.

Our mathematical approach is to transform the learning of a discontinuous field---a known challenge case for neural fields---into the learning of an \emph{everywhere continuous} function---a fundamentally easier learning task. In a nutshell, we augment the input coordinates of the neural field with the generalized winding number of any given cut line~\cite{Jacobson:2018:GWN}, effectively lifting the input from two to three dimensions. Lifting creates 3D distance between corresponding sides of the cut, allowing the network to focus on the easy learning of a \emph{continuous} volumetric field. A corresponding restriction operator enables final output \emph{precisely} resolving \emph{strict} discontinuities, even for cuts unseen during training.

Highlighting the key ingredients of \emph{winding} and \emph{lifting}, we call this \emph{Wind Lifter.} After training a volumetric continuous neural field once, we can (re)compute the winding number at runtime, without any training, as a cut line is lengthened, moved, or reshaped. By combining the continuous neural field and the winding graph, we generate a field with precise, strict discontinuities, while maintaining the speed of ROM.



With Wind Lifter, the discontinuity placement is decoupled from the neural network's weights, and editing the cut geometry is straightforward. Such editing is not possible with traditional neural fields, where the discontinuities are embedded within the network's weights. As a result, our method enables novel applications, such as user-interactive design of cut shapes. 


In summary, we:
\begin{itemize}
    \item introduce Wind Lifter, a generalizable, precise, and easily editable representation of discontinuities in 2D neural fields; 
    \item integrate this new capability with a discretization-agnostic ROM method to enable reduced simulation of cutting;
    \item demonstrate simulation results with complex cuts and cut shape generalization unparalleled by prior reduced modeling methods, including an interactive design prototype and a real-world comparison.
\end{itemize}


\section{Related Work}

\subsection{Simulating Cutting of Deformable Bodies}

A common approach towards simulating cutting is refining and adapting the mesh to capture discontinuities. This approach needs to balance various considerations, such as the number of newly generated elements \cite{Bielser:2000:ISSC, Mor:2000:MSTM}, the quality of the cut details \cite{Busaryev:2013:AFS}, and material information \cite{Chen:2014:refine}. While this approach is intuitive to model, practical implementation is challenging. Remeshing can \Revision{be} challenging to implement \cite{zhang2019amrex} and computationally expensive \cite{obiols2023adarnet}. By contrast, our approach takes a mesh-free approach through neural fields and completely bypasses the need to remesh.

Hybrid Lagrangian-Eulerian methods like the material point method \cite{jiang2016material} handle cutting without remeshing by employing a combination of grid and particle representations. However, the particle-based representation lacks precise surface definition, resulting in artifacts such as numerical fractures and imprecise cutting \cite{su2022ulmpm,fan2025hybrid}.


Extended finite element methods (XFEM) \cite{Moes:1999:XFEM, Kaufmann:2019:enrich, Koschier:2017:RobustXFEM, xfemCutting2024} maintain a fixed mesh and resolve discontinuities using discontinuous ``enrichment functions.'' Our method also resolves  discontinuities using a discontinuous basis. However, unlike XFEM, our approach does not depend on a background mesh, separating the kinematic and geometric representations. This separation enables the simulation of shapes with intricate thin details (see Figure \ref{fig:teaser}) and multiple cuts intersecting localized region corresponding to one mesh triangle or grid cell, situations that can challenge XFEM. Moreover, our separation allows for fast, reduced-order modeling, whereas XFEM solves partial differential equations (PDEs) in maximal degrees of freedom. 

\subsection{Neural Methods for Simulation}

Neural networks have demonstrated efficiency in various areas of physics-based simulation, including deformable simulation \citep{CHEN:CROM-MPM:2023, Feng:2024:NAH, lyu2024accelerate}, fluid simulation \citep{kim2019deep, deng2023neural, 10.1145/3641519.3657438, tao2024neural}, and collision modeling and handling \citep{Romero:LCCHSD:2021, yang2020learning, Cai:2022:CSDF}, among others.

Among all neural methods for physics-based simulation, our approach aligns most closely with reduced-space simulation techniques, which accelerate simulations by finding a reduced basis. These bases can either be mesh-dependent \citep{Fulton:LSD:2018, shen2021high} or mesh-independent \citep{chang:2023:licrom}, and can be data-based (derived from simulation sequences) \citep{chen2023crom, zong2023neural} or data-free (derived from domain geometry) \citep{Sharp:2023:datafree, Modi:2024:Simplicits}. 

However, \emph{none of the data-free methods accommodate cutting during simulation}. Of the data-based methods, \citet{chang:2023:licrom} is able to reproduce only the cut seen during training, that cut being encoded in the neural network weights. By contrast, our method supports procedural cuts without requiring simulation snapshots for training.

For instance, when the reduced basis is computed without simulation data \citep{Modi:2024:Simplicits}, the training process depends solely on sampling the domain. This approach fails to distinguish between an undamaged shape and one with a zero-volume cut, rendering it incapable of representing discontinuities in data-free training scenarios. To address these limitations, our method expressly represents lines of discontinuity and introduces a function representation capable of capturing evolving cut geometries even for zero-volume cuts.

In the context of discontinuity modeling with neural fields \cite{liu20242d}, \citet{Belhe:2023:DiscontinuityAwareNeuralFields} align feature fields from a triangle mesh to discontinuities, which is effective for compressing 2D physics simulation data. However, their approach learns a feature field defined on a mesh, requiring (re)meshing of the domain if discontinuities are placed, extended, or moved. This dependency makes their method unsuitable for modeling a family of discontinuities necessary for progressive cutting, or allowing cut placement and geometry to be edited interactively. Our approach allows for this broader functionality by separating the representation of the continuous field and the discontinuity.

\subsection{Inside/Outside Descriptors}
\zhecheng{given the new name for this related work, we probably need a bit more of context. In fact now I think about it, I feel like the only representation that can do a "jump in value at boundary", "support represent partial cut state" is GWN, is this true?}
Modelling discontinuities inherently requires querying whether a point is on
one side or the other. Signed distance fields (SDFs) and occupancy functions \cite{mescheder2019occupancy} both come to mind. Both rely on a level set to represent a boundary. However, these approaches represent insideness \zhecheng{interior?} only for closed domains \zhecheng{because they essentially only learn the surface levelset in the ambient space, so it means we have to threshold a volumetric function to get the true surface levelset, then we can get an indicator function that is \emph{sharp/jump condition}. because SDf and occupancy function logits are still continuous and smooth cross the 0 levelset}, whereas a cut may be partial or incomplete, i.e., a curve of discontinuity need not be a closed curve. Indeed, an open curve of discontinuity may evolve in time to close up, further begging for a richer representation. 

The winding number is a fundamental concept in mathematics, particularly in complex analysis, geometry, and topology. It measures how many times a closed curve winds around a given point in the plane, by integrating the ``angle of projection'' along the curve with relation to the query point. As a field over all points on the plane it is piecewise constant with jump conditions across the curve~\cite{Reinhart1960TheWN, 10.1145/368637.368653}, making it a natural candidate for ``strict'' insideness tests. \citet{Jacobson:2018:GWN} proposed its generalization (GWN) to open curves, where \Revision{it} is harmonic with jump conditions across the curve, thereby serving as a  ``soft'' insideness test. The GWN has been implemented for point clouds \cite{Barill:FW:2018} and curved surfaces \cite{Feng:2023:WND}, and applied to garment modeling \cite{Hu:2018:TMW, chi2021garmentnets}, geometry processing \cite{Rui2023GCNO, Zhou:2016:MASG}, and computer vision \cite{Mueller:CVPR:2021}. 
Our work leverages the GWN to represent discontinuities. 



\section{How to represent a discontinuous field using a continuous neural network}
\label{sec:lifting}

We aim to accurately represent 2D fields with lines of discontinuity. The field value differs depending on the side from which the discontinuity is approached. Let $f : \Omega \rightarrow \mathbb{R}$ be a real-valued function over $\Omega \subset \mathbb{R}^2$ discontinuous across a curve $\Gamma$, 
\begin{align}
\lim_{\mathbf{x} \to \bm{x_0}^+} f(\mathbf{x}) \neq \lim_{\mathbf{x} \to \bm{x_0}^-} f(\mathbf{x}) \ , \quad \bm{x}_0 \in \Gamma \ , \quad \bm{x} \in \Omega \ ,
\end{align}
\begin{wrapfigure}[6]{r}[-20pt]{0.2\linewidth}
\vspace{-2ex}
\includegraphics[width=1.2\linewidth]{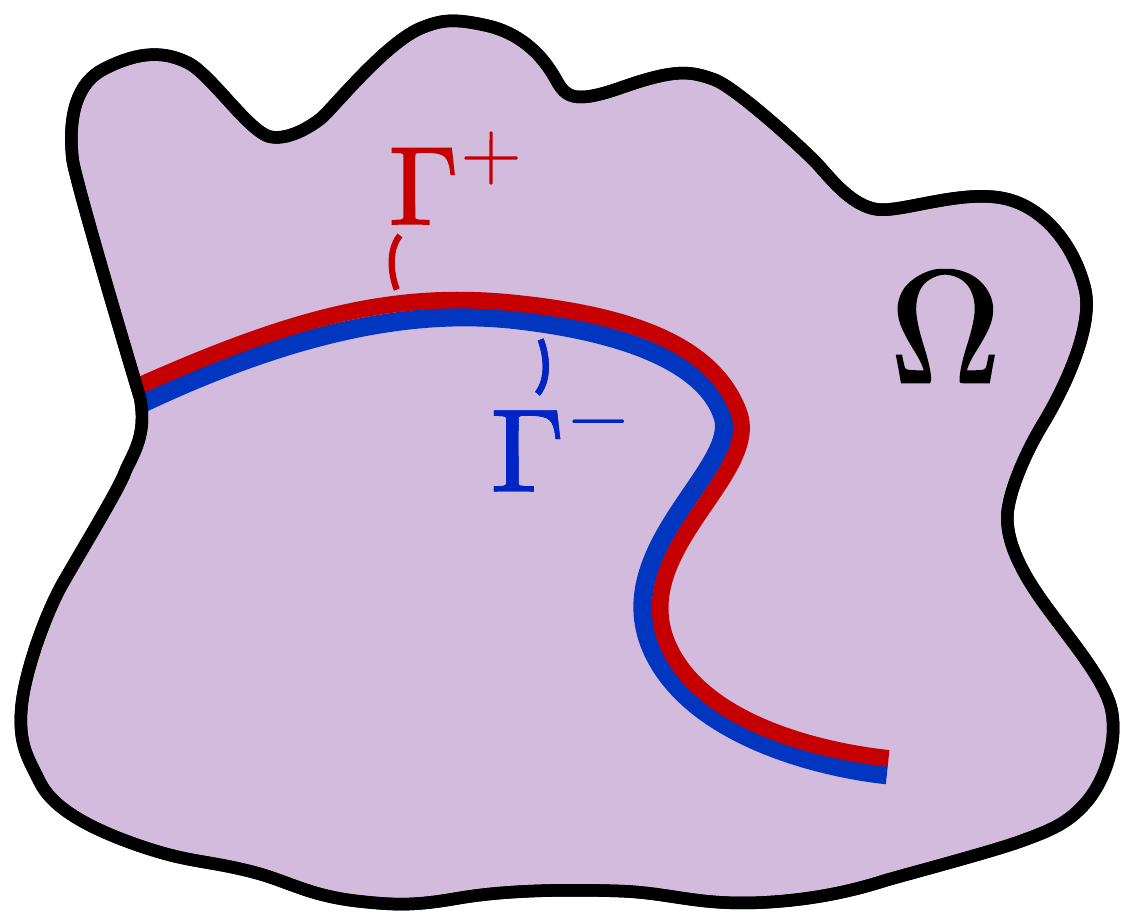}
\end{wrapfigure}
where $\bm{x_0}^+$ and $\bm{x_0}^-$ indicate approaching $\bm{x_0}$ from opposing sides 
$\Gamma^+$ and $\Gamma^-$, respectively (see incident figure).
With a discontinuity, arbitrarily small changes to the evaluation point may produce a finite jump in the field value. We would like to leverage available tools for training neural fields to learn such a function. Unfortunately, typical neural field architectures are poorly suited for this representation task.

\paragraph{Challenges}
Neural fields are typically continuous and differentiable because they are represented by neural networks, such as fully connected feedforward networks, which use smooth activation functions (e.g., sigmoid, or sine). \citet{Rahaman:2019:Spectral} study the spectral bias of such networks, observing that smooth activation functions naturally favor smoother outputs, and common architectures have a natural tendency to learn low-frequency components of a target function first. These properties pose a challenge to precisely representing sharp changes or discontinuities. 

Some techniques may ameliorate, but do not inherently resolve this challenge. For instance, input encoding methods, such as Fourier feature mappings or sinusoidal positional embeddings~\cite{Tancik:2020:Fourier}, and improved activation functions, such as SIREN~\cite{Sitzmann:2020:Implicit}, improve the ability to approximate high-frequency details, but they do not inherently and explicitly model strict discontinuities. 

\paragraph{Proposed architecture}
We do \emph{not} seek to improve what appears to be a fundamental inability of neural fields to represent strict discontinuities. We avoid this limitation and embrace the low-frequency bias of neural fields. We will use feature augmentation to make the learning task easier, simplifying our task to learning only a \emph{continuous} field.

Let $\Tilde{f}_{\theta} : \Omega \times \mathbb{R} \rightarrow \mathbb{R} : (x,y,z) \mapsto \Tilde{f}_{\theta}(x,y,z)$ be a neural field over a \emph{volumetric} domain $\Omega \times \mathbb{R} \subset \mathbb{R}^3$ parameterized
by neural network weights $\theta$. Our learning task will seek a \emph{smooth} neural field $\Tilde{f}_{\theta}$, without discontinuities, by appropriately choosing and leveraging the augmented feature $z$.

Let $z=H(\bm{x})$, where $\bm{x}=(x,y)$. Here, $H: \Omega \rightarrow \mathbb{R}$ is a function that is discontinuous over $\Gamma$. We require that $H$ have an analytical representation, that is, $H$ need not be trained using a neural network. 

\begin{wrapfigure}[4]{r}[-10pt]{0.25\linewidth}
\vspace{-5ex}
\includegraphics[width=1.2\linewidth]{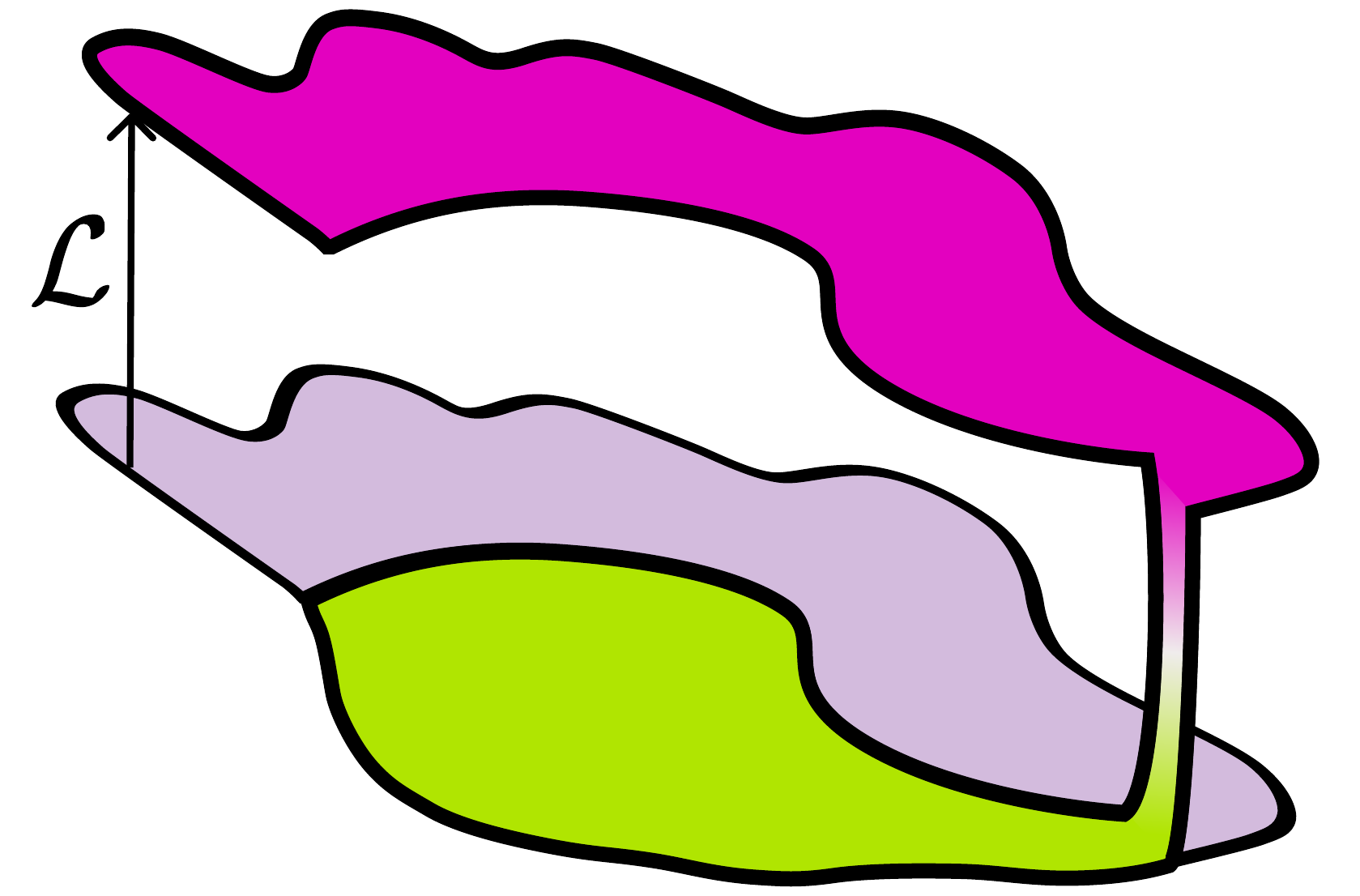}
\end{wrapfigure}
\paragraph{Lifting}
Our augmentation has a simple geometric interpretation. As depicted in the incident figure, our approach lifts the domain into a graph of $H(\bm{x})$ over $\Omega$:
\begin{align}
    \mathcal{L}(\bm{x}) = (\bm{x},H(\bm{x})) = (x,y,H(x,y)) \ , \quad \bm{x} \in \Omega \ .
\end{align}
Since $H$ is discontinuous over $\Gamma$, the graph $\mathcal{L}$ has separated boundaries $\Gamma^+$ and $\Gamma^-$.

\paragraph{Central idea: Discontinuity by restriction}
Our central idea is to define the 2D field $f$ as the restriction of the volumetric neural field $\Tilde{f}_{\theta}$ to the graph $\mathcal{L}$:
\begin{align}
\label{eq:restriction}
\boxed{
    f = \Tilde{f}_{\theta}\circ \mathcal{L} \ , \quad \textrm{or} \quad
    f(\bm{x}) = \Tilde{f}_{\theta}(\mathcal{L}(\bm{x})) \ ,\quad \bm{x} \in \Omega \ .
    }
\end{align}



\paragraph{Why it works.} 
The benefit of lifting is that it decouples spatial discontinuity from the trainable continuous function $\Tilde{f}_{\theta}$. Without lifting, the jumps across $\Gamma$ must be captured by the weight of the neural field, which complicates training. However, our method alleviates this problem. By lifting, two positions $ \bm{x}_0^- \to \Gamma^- $ and $ \bm{x}_0^+ \to \Gamma^+ $, which are close in $\Omega $, are mapped to distinct coordinates in 3D, enabling $\Tilde{f}_{\theta}$ to represent a smooth function, while the discontinuity emerges from its restriction to the discontinuous graph.

As a result, optimizing network weights becomes easier, as no sharp jumps need to be fit, as shown in the very right part in \reffig{fig:hand}. Another advantage of this construction is that it simplifies editing of the field $f$, as the output can be directly controlled by modifying the height function $H(\mathbf{x})$, as depicted in Figure \ref{fig:Interaction}.


\paragraph{Winding graph}
While any function discontinuous over $\Gamma$ may serve as the height $H$, we choose the generalized winding number field~\cite{Jacobson:2018:GWN}, a harmonic function with jump boundary condition across $\Gamma$ amenable to fast evaluation~\cite{Barill:FW:2018} 
\begin{align}
\label{eq:fixedUpper}
    H(\mathbf{x}) = \int_{0}^{1} \frac{\mathbf{\Gamma}'(s) \cdot (\mathbf{\Gamma}(s) - \mathbf{x})^\perp}{|\mathbf{\Gamma}(s) - \mathbf{x}|^2} \, ds \ ,
\end{align}
where $\Gamma : (0,1) \rightarrow \Omega : s \mapsto \Gamma(s)$, and
$(\cdot)^\perp$ rotates a vector by $\pi/2$. We refer to the graph of $H(\mathbf{x})$ as the \emph{winding graph}.

\paragraph{Progressive cutting} 
Compared to a permanent discontinuity, progressive cutting involves the notion that the curve of discontinuity lengthens over time. We introduce the parameter $\alpha \in [0,1]$ to mark the fraction of $\Gamma$ that has been cut thus far. We use $\Gamma^\alpha$ to denote the portion of $\Gamma$ corresponding to the first $\alpha$-fraction of its length, and we account for $\alpha$ by modifying the integration bounds:
\begin{align}
\label{eq:winding_para}
    H^\alpha(\mathbf{x}) = \int_{0}^{\alpha} \frac{\mathbf{\Gamma}'(s) \cdot (\mathbf{\Gamma}(s) - \mathbf{x})^\perp}{|\mathbf{\Gamma}(s) - \mathbf{x}|^2} \, ds \ .
\end{align}
Similarly, since the cut progression generally affects the optimal kinematic basis, we also condition the volumetric neural field on $\alpha$:
\begin{align} \label{eq:neural_para}
    \Tilde{f}^\alpha_\theta : [0,1] \times \Omega \times \mathbb{R} \rightarrow \mathbb{R} : (\alpha,x,y,z) \xmapsto{\textrm{\scriptsize inference}}   \Tilde{f}^\alpha_{\theta}(x,y,z)  \ .
\end{align}
The lifting and restriction operators are unaffected, but for completeness we indicate their dependency on $\alpha$:
\begin{align} \label{eq:lifting_para}
    \mathcal{L}^\alpha(\bm{x}) = (\bm{x},H^\alpha(\bm{x}))  \quad \textrm{and}\quad
    f^\alpha(\bm{x}) = \Tilde{f}^\alpha_{\theta}(\mathcal{L}^\alpha(\bm{x})) \ .
\end{align}

\paragraph{Outlook}
Equations \ref{eq:winding_para}--\ref{eq:lifting_para} summarize the construction of a  function $f^\alpha : \Omega\rightarrow \mathbb{R}$ discontinuous across part of the curve $\Gamma$ depending on the cutting extent $\alpha$. As we have seen, $f^\alpha$ is constructed by restricting a learned volumetric neural field ($\Tilde{f}_\theta)$ to the \emph{winding graph} $H(\mathbf{x})$ using the \emph{lifting operator} $\mathcal{L}$.

We are ready to construct ROMs that precisely resolve discontinuities, applying the same constructions and notation to kinematics-specific fields.

\section{Reduced Order Modeling}

\paragraph{Kinematics} Consider a thin-walled elastic body parameterized by the domain $\Omega \subset \mathbb{R}^2$. The deformed position of the body in three-space is given by the displacement field $\bm{u}(\bm{x}): \Omega \rightarrow \mathbb{R}^3$. Reduced order modeling (ROM) seeks to represent this displacement field using a small number $k$ of coordinates, that is, via a reduced configuration $\bm{z} \in \mathbb{R}^k$~\cite{An:Cubature:2008,Kim:Skipping:2009,barbivc2005real}. In particular, linear ROM, which we will consider here, requires that $\bm{u}$ be \emph{linear} in $\bm{z}$, that is, 
$\bm{u} = \bm{z}^T \bm{\Phi}$, where $\bm{\Phi^T} = [\bm{\phi}_1, \ldots, \bm{\phi}_k]$, and $\bm{\phi}_i : \Omega \rightarrow \mathbb{R}^3$ is a displacement \emph{basis}. 

In our setting, the displacement basis must represent a discontinuity over $\Gamma$ depending on the cutting extent $\alpha$. Therefore, $\bm{\Phi}$ is dependent on $\alpha$. We will be reminded of this dependency with the notation $\bm{\Phi}^\alpha$, and omit the decorations where it is clear from context. 

\subsection{Precomputation: Training}

\paragraph{Training overview}
Recalling Equation \ref{eq:restriction}, we construct a displacement basis field $\bm{\phi}^\alpha$ discontinuous across $\Gamma$ by training a volumetric neural field $\Tilde{\bm{\phi}}_\theta^\alpha$ continuous over $\Omega \times \mathbb{R}$. 
While $\Tilde{\bm{\phi}}_\theta^\alpha$ and $\bm{\phi}^\alpha$ are vector-valued, nothing changes in the lifting construction, which remains simply 
$\bm{\phi}^\alpha = \Tilde{\bm{\phi}}_\theta^\alpha \circ \mathcal{L}^{\alpha}$.

Our main contribution (\Cref{sec:lifting}) is orthogonal to the choice of training scheme. In fact, our neural discontinuity representation is compatible with both data-driven and data-free approaches. Below, we will develop how to incorporate the proposed winding-number-lifting approach in either of the training setups. Regardless of the approach taken, the goal is to train the weights $\theta$.

\paragraph{Data-driven basis learning.} 
In the data-driven setting, we begin by collecting training snapshots, recording each displacement field $\bm{u}_j$ and cut progression $\alpha_j$ at time increment $j$. The neural field  $\Tilde{\bm{\Phi}}^\alpha_\theta$ is then trained by minimizing the reconstruction loss over all the simulation snapshots:
\begin{align}
\mathcal{L}_{\text{data-driven}} = \sum_{j} \left\| \bm{z}_j^T \Tilde{\bm{\Phi}}_\theta  \circ \mathcal{L}  - \bm{u}_j \right\|^2_2 \ ,
\end{align}
where  
$\bm{z}_j \in \mathbb{R}^k$ is the reduced coordinate corresponding to the optimal reconstruction 
at time increment $j$. Note that $\|f\|^2_2 = \int_\Omega f(\bm{x})^2 \,\mathrm{d}\bm{x}$ is the $L_2$ norm on $\Omega$,
which is estimated via uniform stochastic cubatures~\cite{An:Cubature:2008,carlberg2011model}.
For more details on the training process, please refer to \cite{chang:2023:licrom}. The key distinction in our approach is the incorporation of cutting, achieved by restricting the neural field to the lifting function. Furthermore, we ensure that both $\Phi$ and $\mathcal{L}$ explicitly depend on $\alpha_j$, the cut progression parameter.



\paragraph{Data-free basis learning.} 

In addition to the previous data-driven training setting, \Revision{our neural field} can also be trained in a data-free fashion. Following \citet{Modi:2024:Simplicits}, we trained the neural network by minimizing the elastic energy 
\begin{align} \label{eq:training-data-free}
    \mathcal{L}_{\text{data-free}} = E_{\text{elas}} = \int_{\Omega}\Psi(\bm{u}(\bm{x})) \, \mathrm{d}\bm{x} 
\end{align}, where $\Psi$ is the elastic energy density, e.g., St. Venant-Kirchhoff (StVK) material \cite{barbivc2005real}. For more details on the data-free training, please refer to the supplementary material and \citep{Modi:2024:Simplicits}. We emphasize that the loss function only involves an analytically defined elastic energy and does not involve any training data (e.g., from full-order simulations).

\subsection{Dynamic Subspace Simulation}
\label{sec:reduced}
Following \citet{chang:2023:licrom}, the reduced configuration $\bm{z}$ is updated at each time step via the optimization
\begin{align} 
\label{eq:time_stepping}
 \bm{z}_{j+1} = \operatorname{argmin}_{\bm{z}} \frac{1}{2} \|\bm{z} - \bm{z}_{j+1}^{\text{pred}} \|^2 + h^2 \int_{\Omega} \Psi\left( \bm{u}(\Revision{\alpha}, \bm{x}, \bm{z}) \right) \, d\bm{x} \ ,
\end{align}
where $h$ is the time step size, and $\bm{z}_{j+1}^{\text{pred}} = 2\bm{z}_{j} - \bm{z}_{j-1}$. Evaluating the elastic energy $\Psi$ involves computing the deformation gradient $\mathbf{F} = \nicefrac{\partial \bm{u}(\alpha, \bm{x}, \bm{z})}{\partial \bm{x}}$. This gradient can be directly obtained from the reduced-space coordinate $\bm{z}$ and the spatial gradient of the neural field, $\nicefrac{\partial \bm{\phi}_{\bm{x}}^{\alpha}}{\partial \bm{x}}$, which is efficiently calculated via automatic differentiation of the neural network. We evaluate the domain integral using uniform stochastic cubature. Our examples optimize \eqref{eq:time_stepping} using gradient descent; we also implemented Newton's method and observed similar performance.

\subsection{Implementation}

\paragraph{Progressive cutting and cut placement editing}
Our progressive cutting simulations increase the cutting extent $\alpha$ over time. Our interactive design application modifies $\Gamma$ during simulation. Changing either $\alpha$ or $\Gamma$ immediately affects the winding graph, which is computed as-needed analytically; crucially, it does not require retraining the volumetric neural field.

Our implementation represents $\Gamma$ as a collection of polylines (allowing for more than one cut). The evaluation of the generalized winding number locates the appropriate bounds of integration by walking along the polyline until the fractional length $\alpha$ has been walked (refer to supplemental material).

\paragraph{Strain singularity at crack tip}
The elastic energy $\Psi$ involves the spatial gradient of displacements $\nabla_{\bm{x}}\bm{u}$, and in turn $\nabla_{\bm{x}}H$. While the winding number itself is bounded, its gradient diverges approaching the endpoints of $\Gamma$. In mechanics this is known as the crack tip strain singularity, a recognized challenge to force calculations~\cite{Mousavi:XFEM:2011}. \zhecheng{I kinda get it, but has a didactic figure here would be perfect} Following typical treatments, we smooth the $\epsilon$-neighborhood around the endpoints of $\Gamma^\alpha$ by multiplying the generalized winding number by a cubic spline kernel function (refer to supplemental material).

\paragraph{Joint training}
Since $\bm{\Phi}$ is a $k$-dimensional basis, we must repeat the lifting construction $k$ times. Conceptually, the simplest approach is to train $k$ volumetric networks, however, this is not strictly necessary, since a volumetric neural field is not limited to a 1-dimensional (real-valued) output. In our implementation, we parameterize all $k$ displacement bases using the same neural network weights $\theta$ by making $\Tilde{\bm{\phi}}_\theta^\alpha$ a $k$-valued field, i.e.,
$\bm{\phi}_i^\alpha = \Tilde{\bm{\phi}}_{\theta,i}^\alpha \circ \mathcal{L}^{\alpha}$.

\section{Results}

Table \ref{tab:timing_table} summarizes the timing statistics for the examples in our paper. Time per step increases with cut changes due to the need to re-evaluate the neural network and update the basis for the new cut shape. Without cut changes, our method matches traditional reduced space simulations. Our method runs at $28.5 \sim 42.4$ fps with cut changes and $62 \sim 166$ fps without. For reference, full-space simulation takes around 1.1s per frame without cut changes during simulation. This gives our method a $30\times \sim 183\times$ speedup over full-space simulations while enabling cut changes during simulation. All results are reported using an NVIDIA RTX 4090 GPU and an AMD 7950X CPU. \Revision{Unless otherwise noted, we use a 5-layer, 128-channel SIREN MLP with positional encoding up to maximum frequency of $2^3$. Winding numbers are scaled by a factor of 32 to ensure sufficient separation of lifted and 2D coordinates.}

\paragraph{Discretization-agnostic domain representation} Our method does not assume a specific type of domain discretization, instead relying only on the ability to sample the domain. All examples below use stochastic cubature drawn from a uniform distribution over the domain~\cite{chang:2023:licrom,Modi:2024:Simplicits}. Figures \ref{fig:kirigami_chain} and \ref{fig:Comparison_results} are rendered with a mesh \Revision{bound} to the shape, while other figures are visualized as point clouds with colors sampled from a texture.

\subsection{Evaluation on Data-Free Settings}

\begin{figure} 
\centering
\includegraphics[width = 0.9\linewidth]{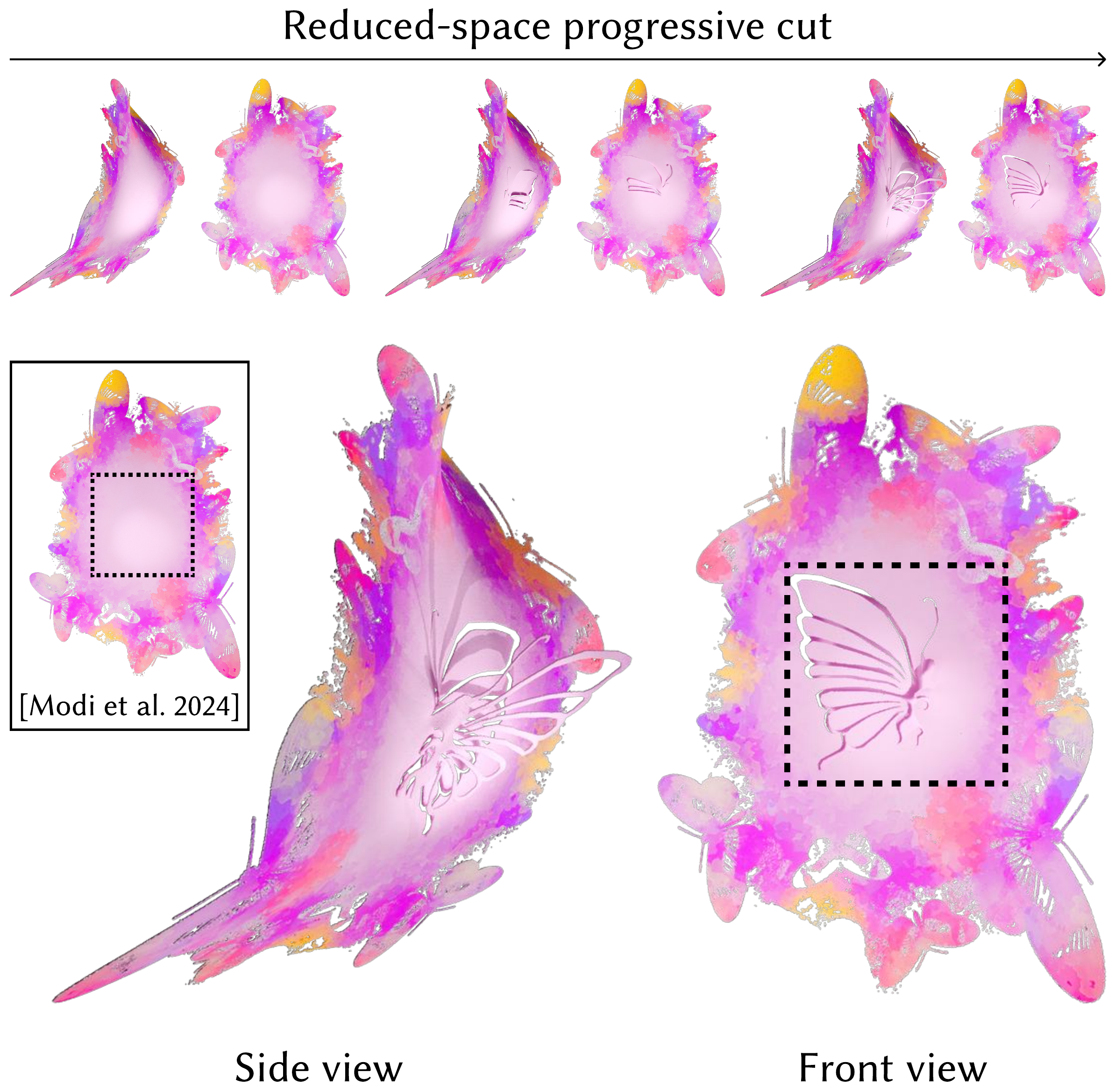}


\caption{Our method is able to capture the complex cut of this butterfly shape, this is not achievable by previous data-free method \cite{Modi:2024:Simplicits} that uses traditional MLP due to the built-in continuity.}
\label{fig:butterly} 
\end{figure}

\paragraph{Comparison to prior work}
To the best of our knowledge, our method is the first data-free model reduction method to support displacement field discontinuities, \Revision{progressive} cutting, or generalization of cut placement. As depicted by the butterfly in Figure~\ref{fig:butterly} and the small ginkgo leaves in Figure~\ref{fig:teaser}, our method enables a ROM with highly detailed cut structures. All of our data-free examples use the $18$-dimensional basis presented above.





\begin{figure}
\centering
\includegraphics[width = \linewidth]{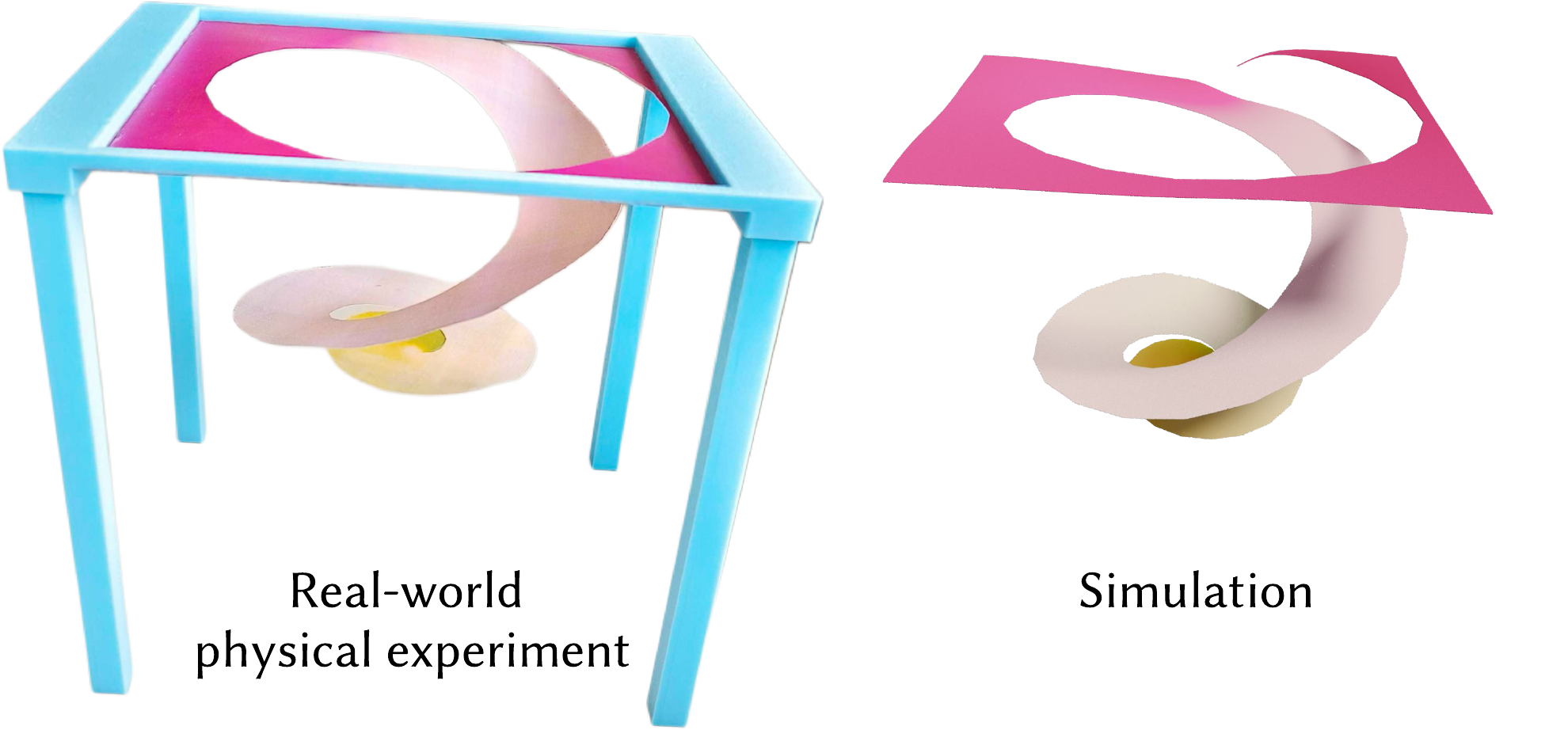}
 \vspace{-2em}
\caption{
Real-world Physical Experiment. We performed a qualitative comparison with a real-world experiment. A helical shape was cut and photographed as it sagged under gravity under the same boundary conditions. The simulated deformation qualitatively matches the real-world data.
} 
\label{fig:realworld} 
\end{figure}

\begin{figure}
\centering
\begin{minipage}[c]{1.\linewidth} 
\begin{tikzpicture}

    \node (img1) at (0\linewidth, 0) {\includegraphics[width=0.26\linewidth]{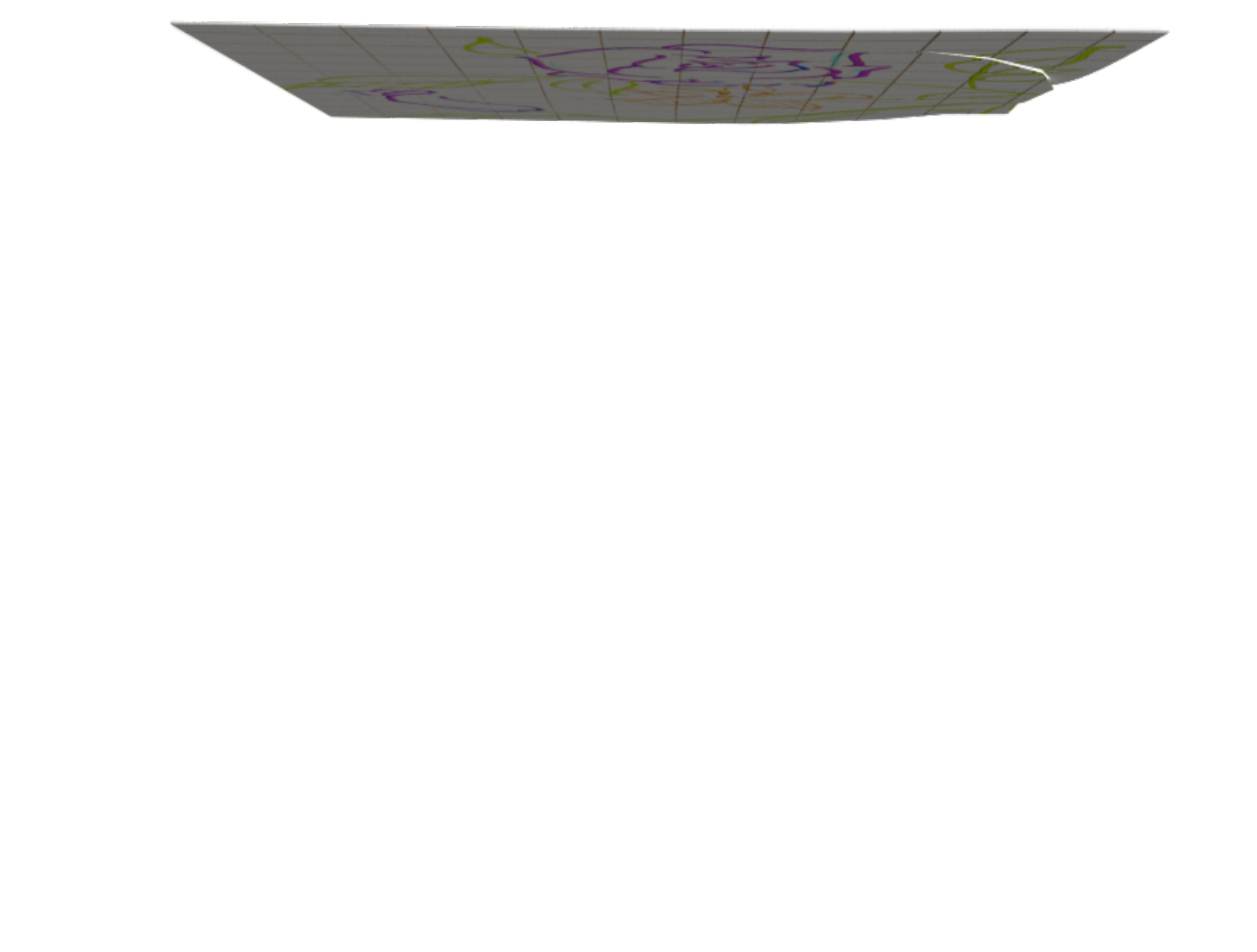}};
    \node (img2) at (0.2366\linewidth, 0) {\includegraphics[width=0.26\linewidth]{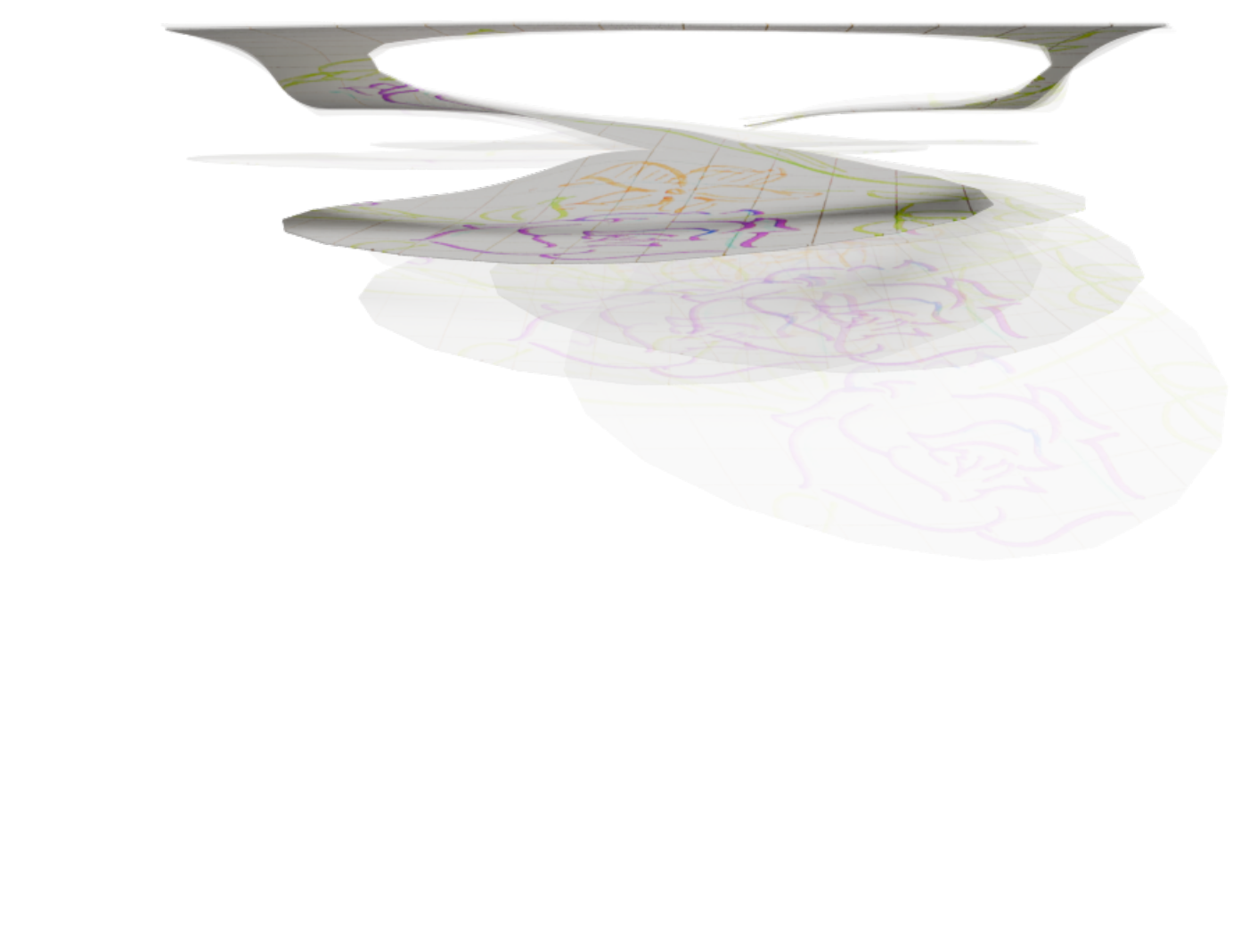}};
    \node (img3) at (0.4732\linewidth, 0) {\includegraphics[width=0.26\linewidth]{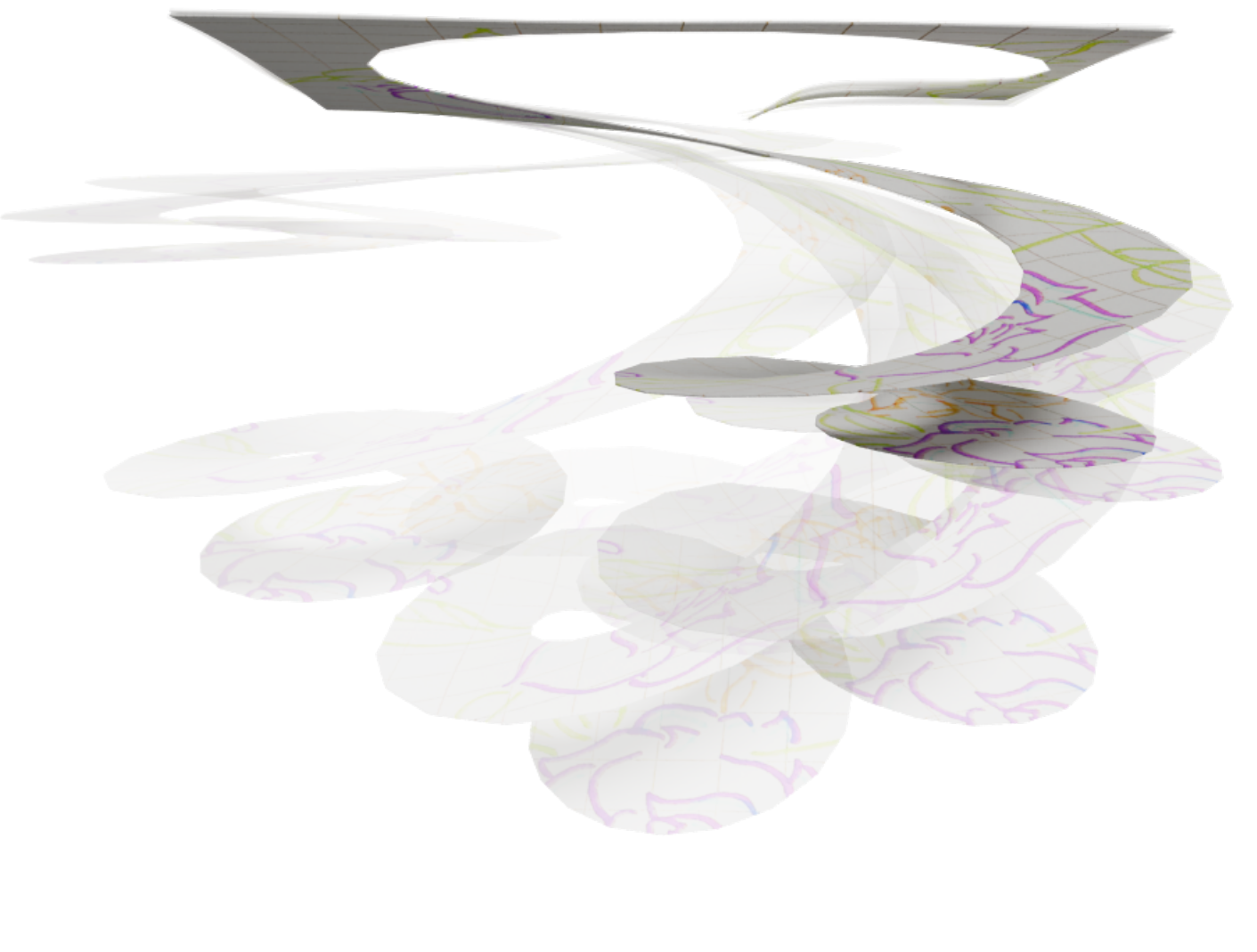}};
    \node (img4) at (0.7098\linewidth, 0) {\includegraphics[width=0.26\linewidth]{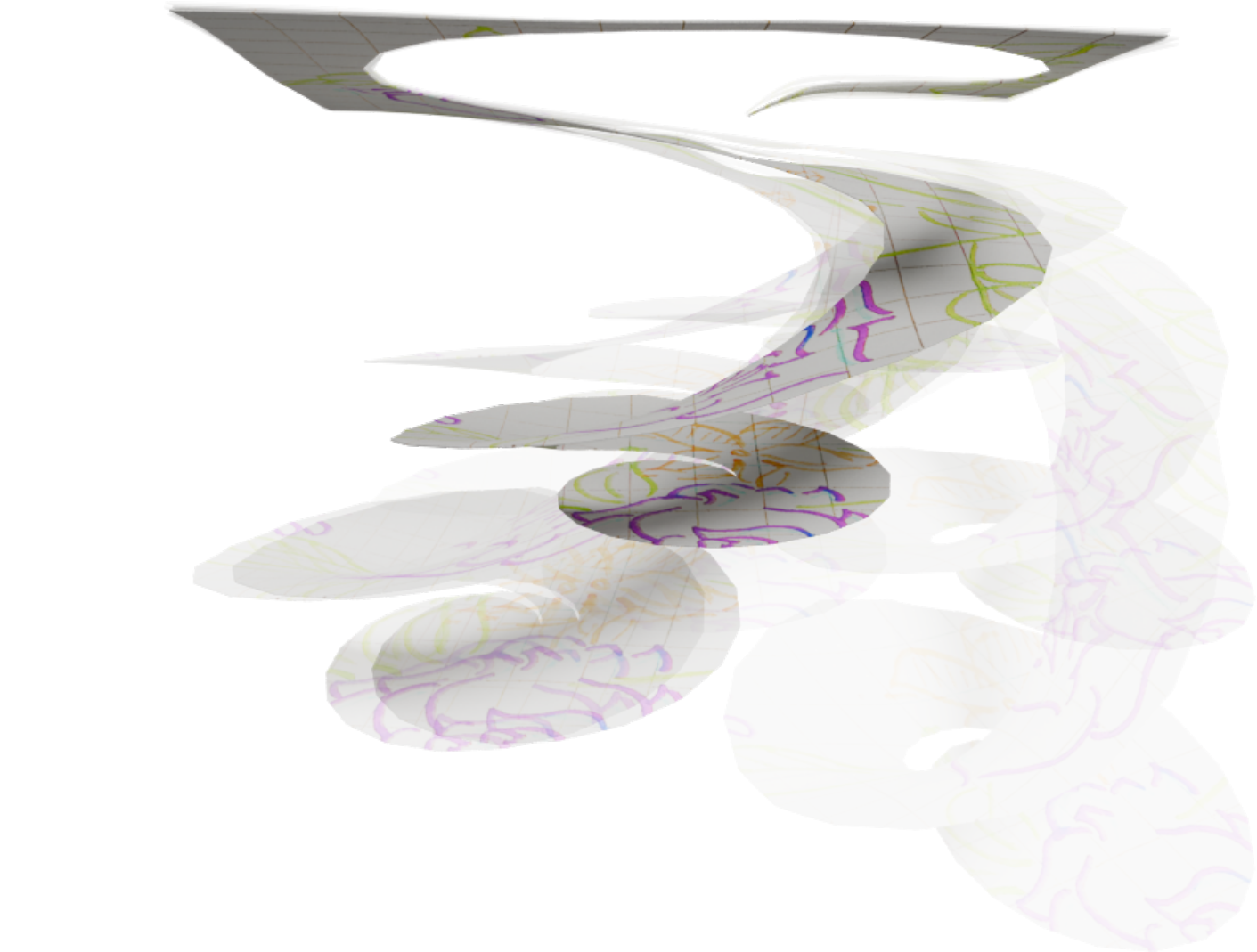}};

    \node[align=center] at ([yshift=1mm]img1.north) {\footnotesize Poke};
    \node[align=center] at ([yshift=1mm]img2.north) {\footnotesize Poke};
    \node[align=center] at ([yshift=1mm]img3.north) {\footnotesize Poke};
    \node[align=center] at ([yshift=1mm]img4.north) {\footnotesize Poke};
    \draw[thin, -{Latex[length=1.25mm]}] 
        ([shift={(0.7,0.8)}]img1.south) 
        to[bend right=40] 
        node[midway, below] {\footnotesize Cut} 
        ([shift={(-0.7,0.8)}]img2.south);
    \draw[thin, -{Latex[length=1.25mm]}] 
        ([shift={(0.7,0.35)}]img2.south) 
        to[bend right=40] 
        node[midway, below] {\footnotesize Cut} 
        ([shift={(-0.7,0.35)}]img3.south);

\end{tikzpicture}
\end{minipage}
 \vspace{-1em}
\caption{
Our method generalizes to loadings (forces) unseen during training. To demonstrate this, we applied a previously unseen force at various stages of the progressive cut, resulting in deformations that reflect the evolving geometry at each stage. } 
\label{fig:GenLoading} 
\end{figure}

\paragraph{Generalization}
We explore generalization to 
\emph{external loads} and \emph{cut geometry} unseen during training.
Since data-free ROMS do not explicitly consider external loads during training, any external force may demonstrate generalization. We applied a sequence of external forces at different stages of a progressive helical cut, inducing varying deformations responsive to the changing geometry of the evolving discontinuity (see Figure \ref{fig:GenLoading} and accompanying video). We cut a helix from real paper and observed that its final sagged state agrees with the simulated result (see Figure \ref{fig:realworld}). 

\begin{figure}
\begin{subcolumns}[0.99\linewidth]
  \subfloat[Training shapes]{\includegraphics[width=\subcolumnwidth]{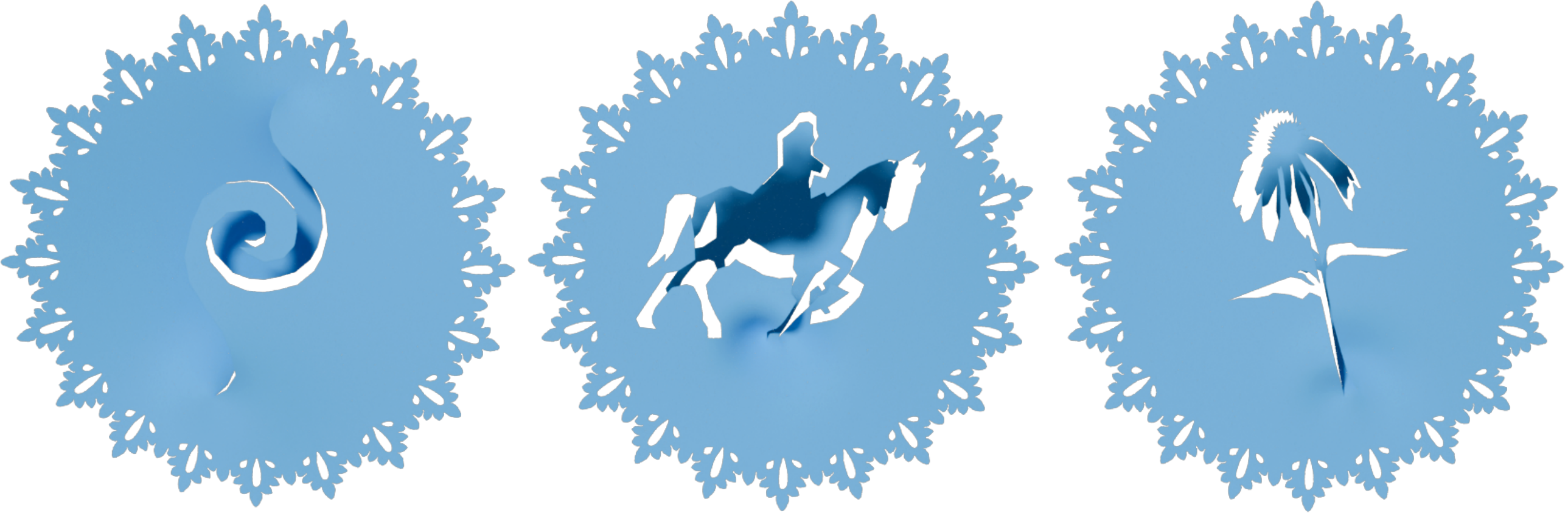}}
   \vspace{-0.5em}
  \subfloat[Testing shapes, not in training data]{\includegraphics[width=\subcolumnwidth]{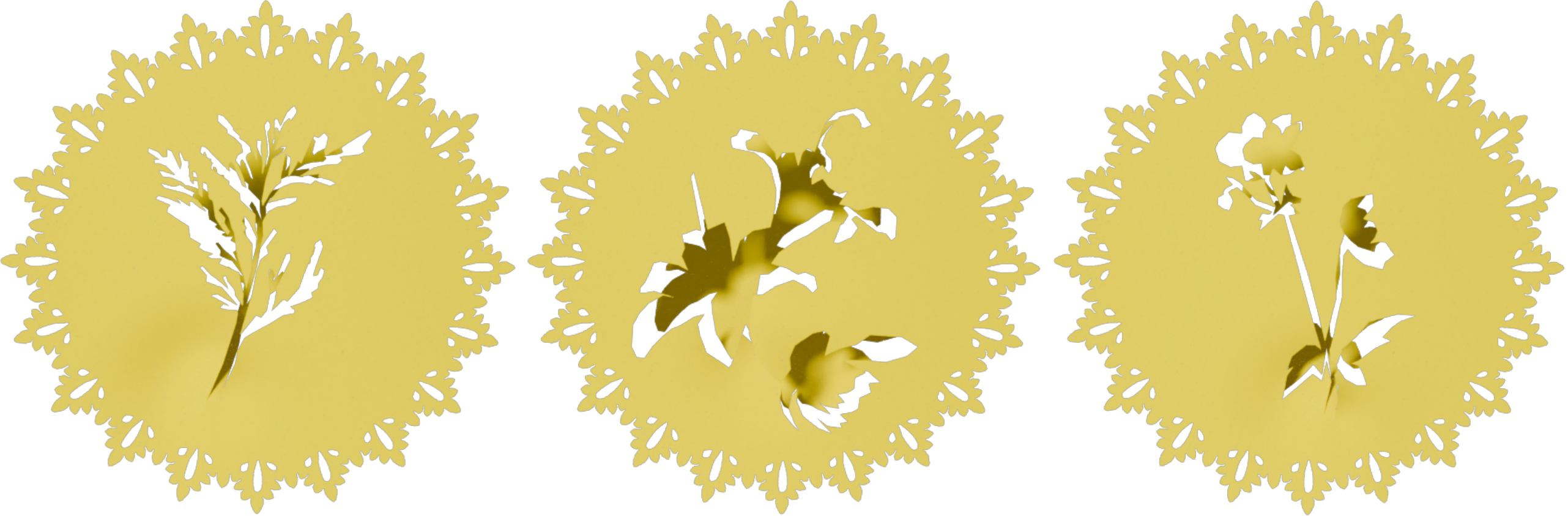}}
\end{subcolumns}
 \vspace{-0.5em}
\caption{We train our method with the three blue cut shapes (top), and test it on three significantly different cut shapes not included in training (bottom). We can still get reasonable deformation for the testing set.}
\label{fig:GenGeo1}
\end{figure}

To evaluate generalization over cut geometry, we trained the network weights $\theta$ on three ``training'' cuts, then froze the weights and simulated three significantly different 'test' cuts, observing plausible results (see Figure~\ref{fig:GenGeo1} and accompanying video).

\begin{figure}
\includegraphics[width = 8cm]{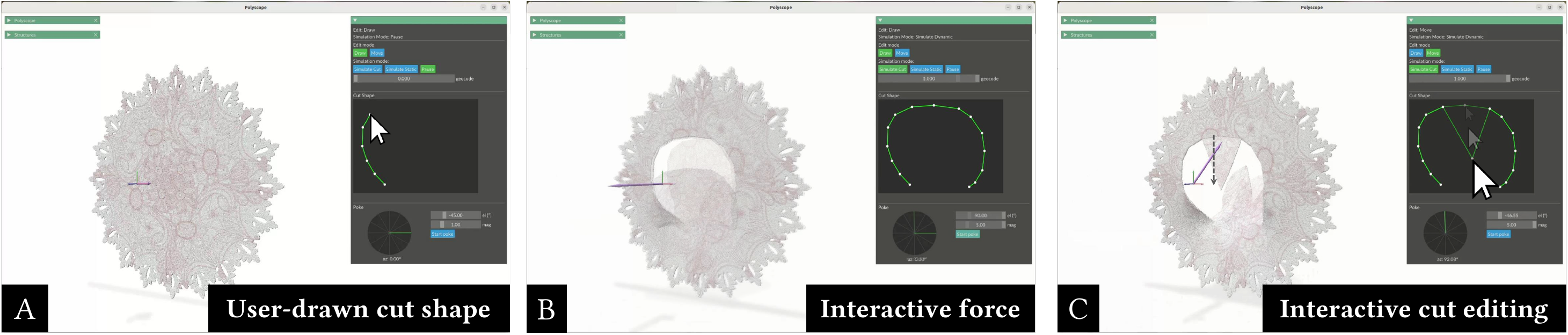}
 \vspace{-0.5em}
\caption{Our method enables interactive drawing and editing of cut shapes with real-time previews. Users can drag polyline vertices to modify the cut shape during an ongoing simulation, in real-time. This is highly challenging for mesh-based methods due to the frequent need to constantly re-mesh the entire domain and has not been demonstrated.}
\label{fig:Interaction}
\end{figure}

\paragraph{Interactive design}
We prototyped an interactive cut editor (see Figure \ref{fig:Interaction}). The designer draws and edits the cut geometry, applies external forces, while observing the corresponding deformation in real time. The cut position and shape may be edited without restarting the simulation, with the deformation updating in real time (refer to accompanying video). To the best of our knowledge, real-time physical preview as a cut geometry is altered has not been reported in the literature. Implementing interactive cut editing with a pure neural field representation~\cite{chang:2023:licrom} is not possible since the discontinuity is ``baked in'' to the network weights; on the other hand, if a mesh-based discontinuity representation is introduced~\cite{Belhe:2023:DiscontinuityAwareNeuralFields}, interactive editing would necessitate ongoing remeshing to align mesh edges with $\Gamma$, which has not been demonstrated, and has the potential to be costly or to exhibit basis-projection artifacts.


\begin{table*}[ht]
\centering
\rowcolors{2}{white}{gray!20} 
\begin{tabularx}{\linewidth}{l |>{\centering\arraybackslash}X |>{\centering\arraybackslash}X |>{\centering\arraybackslash}X |>{\centering\arraybackslash}X}
 & \textbf{Number of Vertices (k)}& \textbf{Training Time (s)} & \makecell{\textbf{Time per Step}\\ \textbf{With Cut (ms)}} & \makecell{\textbf{Time per Step}\\ \textbf{Without Cut (ms)}} \\
\textbf{hand} (\reffig{fig:hand}) & 50 & 53 & 23.56 & 8.57 \\
\textbf{leaf} (\reffig{fig:teaser}) & 50 & 78 & 25.40 & 9.96 \\
\textbf{butterfly} (\reffig{fig:butterly}) & 90 & 73 & 34.86 & 14.89 \\
\textbf{helical} (\reffig{fig:GenLoading})& 100 & 161 & 35.32 & 14.77 \\
\textbf{generalization} (\reffig{fig:GenGeo1}) & 100 & 81 & 35.11 & 15.15 \\
\textbf{kirigami} (\reffig{fig:kirigami_chain})& 137 & 2760 & 30.95 & 16.01 \\
\textbf{slit} (\reffig{fig:Comparison_results})& 20 & 1740 & 25.33 & 6.02 \\
\end{tabularx}
\caption{We provide the number of vertices and timing details for all examples. Thanks to the compact neural representation, our approach significantly reduces training time compared to many neural methods that require hours or even days. Our training time ranges from just under a minute to slightly less than an hour. During inference, the time cost is higher when cut changes are involved, as these require both neural network inference and basis updates.}
\label{tab:timing_table}
\end{table*}



\subsection{Evaluation on Data-Driven Settings}
We also evaluate our method in data-driven settings. The training data and ground-truth full-space simulations for all examples in this section are generated using the data generation code publicly released by~\citet{chang:2023:licrom}, which implements a FEM solver for stable Neo-Hookean energy~\cite{Smith:2018:stablenh}. We train $k=20$ displacement modes for all examples in this section. 

\paragraph{Interactive deformation editing}

\begin{figure}
\centering
\includegraphics[width=\linewidth]{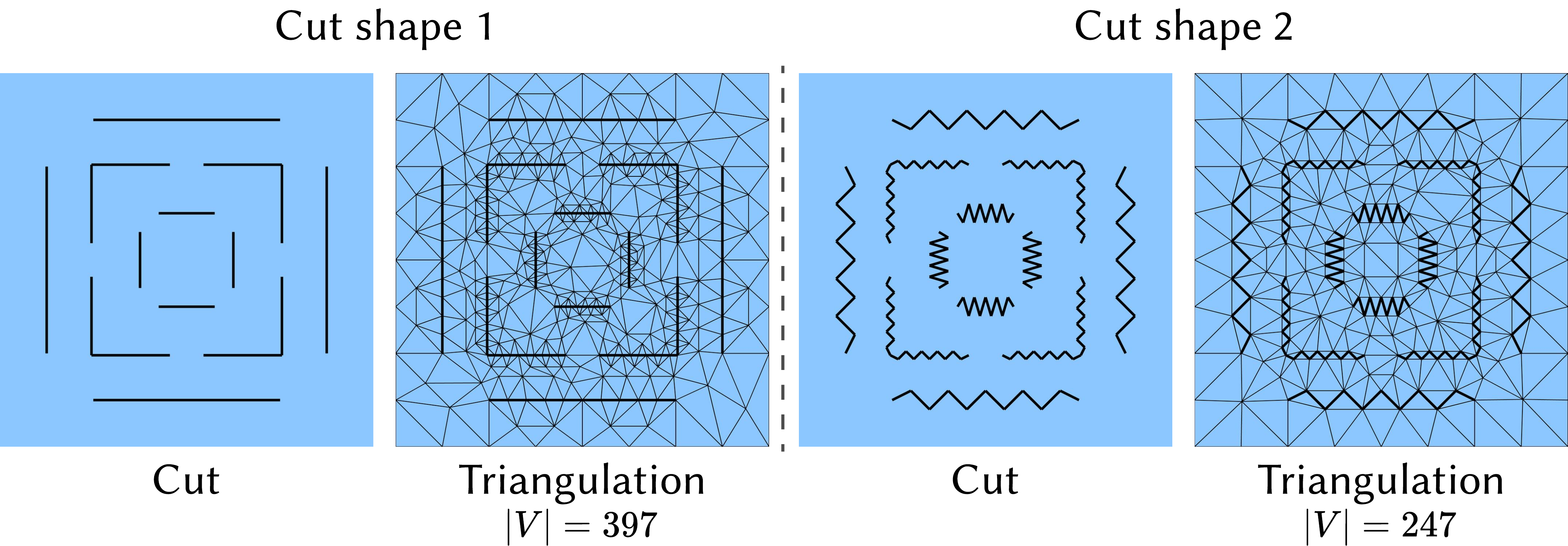}
\caption{The cut change can lead to a big difference in discretization using the mesh-based method. We applied constrained Delaunay triangulation with the same setting for these two cut shapes, and the resulting number of vertices is significantly different.}
\label{fig:remeshing}
\end{figure}

\begin{figure}
\includegraphics[width=8cm]{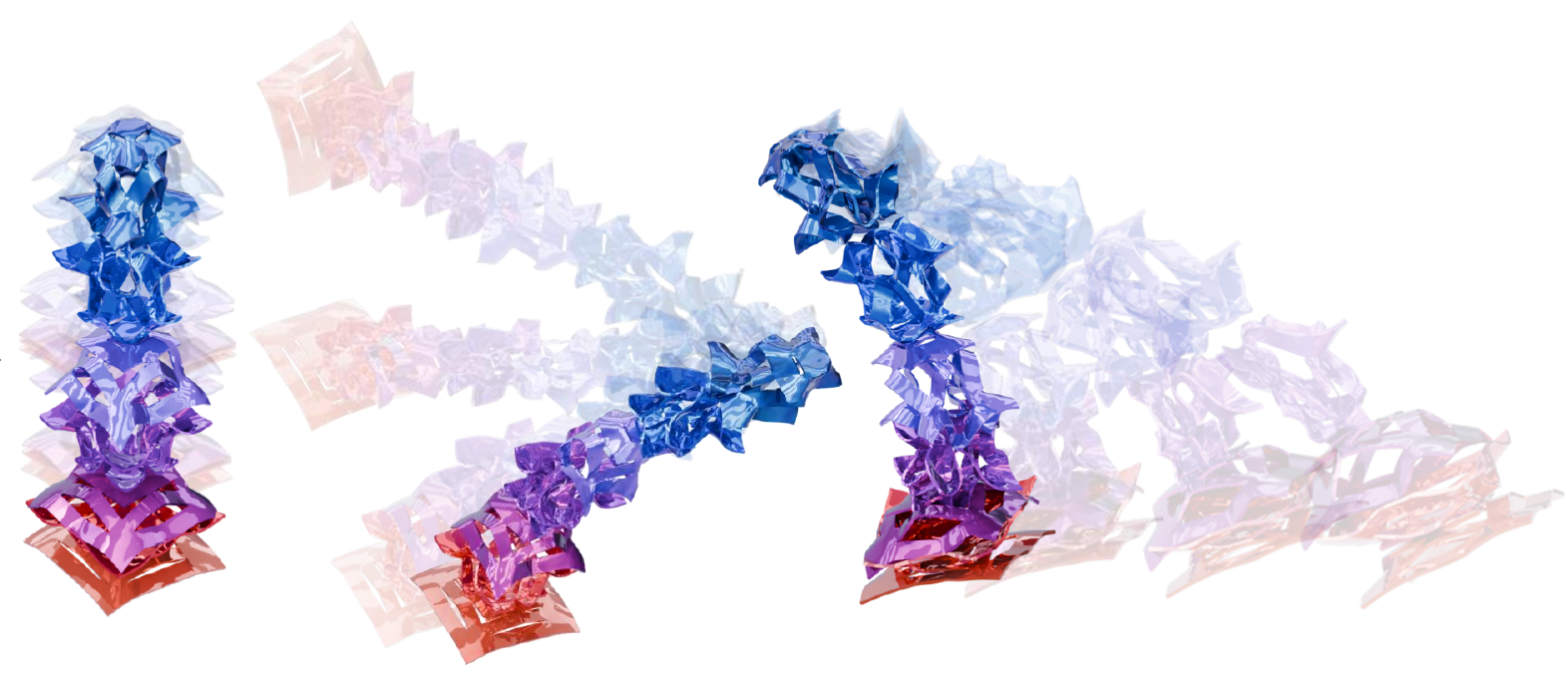}
\caption{We trained the basis on simulations of tugging on a \emph{single} kirigami sheet. By connecting many instances of this kirigami sheet end to end, we obtain a reduced simulation of this ``kirigami tower.''}
\label{fig:kirigami_chain}
\end{figure}

We trained a ROM on simulations of tugging on square sheet with various etched angular cuts (see Figures \ref{fig:kirigami_chain} and \ref{fig:remeshing}). By connecting many instances of this ROM end to end, we obtain a reduced simulation of a ``kirigami tower.''

\begin{figure}
\includegraphics[width=8cm]{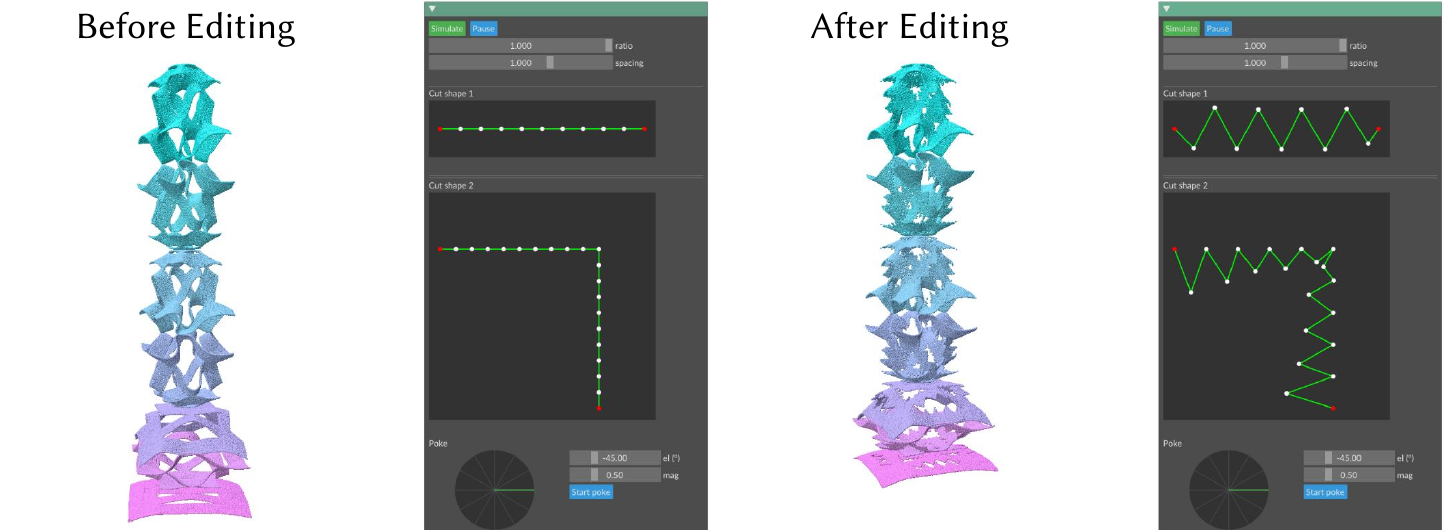}
\caption{We can interactively edit the cut shape for the kirigami. The cut lines of the kirigami change from a straight line to a 'z' shaped curve.}
\label{fig:kirigami_interactive}
\end{figure}

We demonstrate generalizability with an interactive interface that enables users to edit the cut geometry while the simulation is active (see Figures \ref{fig:kirigami_interactive} and \ref{fig:remeshing}). Starting from a straight cut, users can drag and adjust the control points of the kirigami cuts. After editing, the resulting zig-zag cut shape significantly differs from the initial straight cut, which was not included in the training data (Figure \ref{fig:remeshing}). 

This type of editing is particularly challenging for prior mesh-based methods, including neural fields that rely on meshes \cite{Belhe:2023:DiscontinuityAwareNeuralFields}. As shown in Figure \ref{fig:remeshing}, we applied constrained Delaunay triangulation \cite{Richard:2005:triangle} to two different cut shapes. The resulting number of vertices and triangle arrangement for the two cut patterns differs significantly, highlighting that remeshing is a global operation and that alignment of edges and points to revised cut lines would be a non-trivial (unexplored) alternative.

\vspace{-0.1cm}
\paragraph{One-shot generalization}
To evaluate quantitatively the generalizability of our method, we compare it with previous approaches that also use neural fields to construct the basis for reduced-space simulation, including LiCROM \cite{chang:2023:licrom} and Simplicits \cite{Modi:2024:Simplicits}. As a third baseline, the DANN network proposed by \citet{Belhe:2023:DiscontinuityAwareNeuralFields}, which is designed to better capture discontinuities, replaces the standard MLP in LiCROM, keeping the remainder of LiCROM \Revision{intact}.

We first run full-space simulations on cuts at different positions and train all methods using simulation snapshots from one cut position. For Simplicits, which does not require simulation trajectories and only samples the rest shape, we trained solely on the undeformed shape samples found in the same training set.

\begin{figure*} 

\centering
\begin{minipage}[c]{0.24\linewidth}

\begin{tikzpicture}[spy using outlines={rectangle,black,magnification=3,size=2.2cm}]

    \node[transform shape] (image) {\pgfimage[interpolate=true,width=0.95\linewidth]{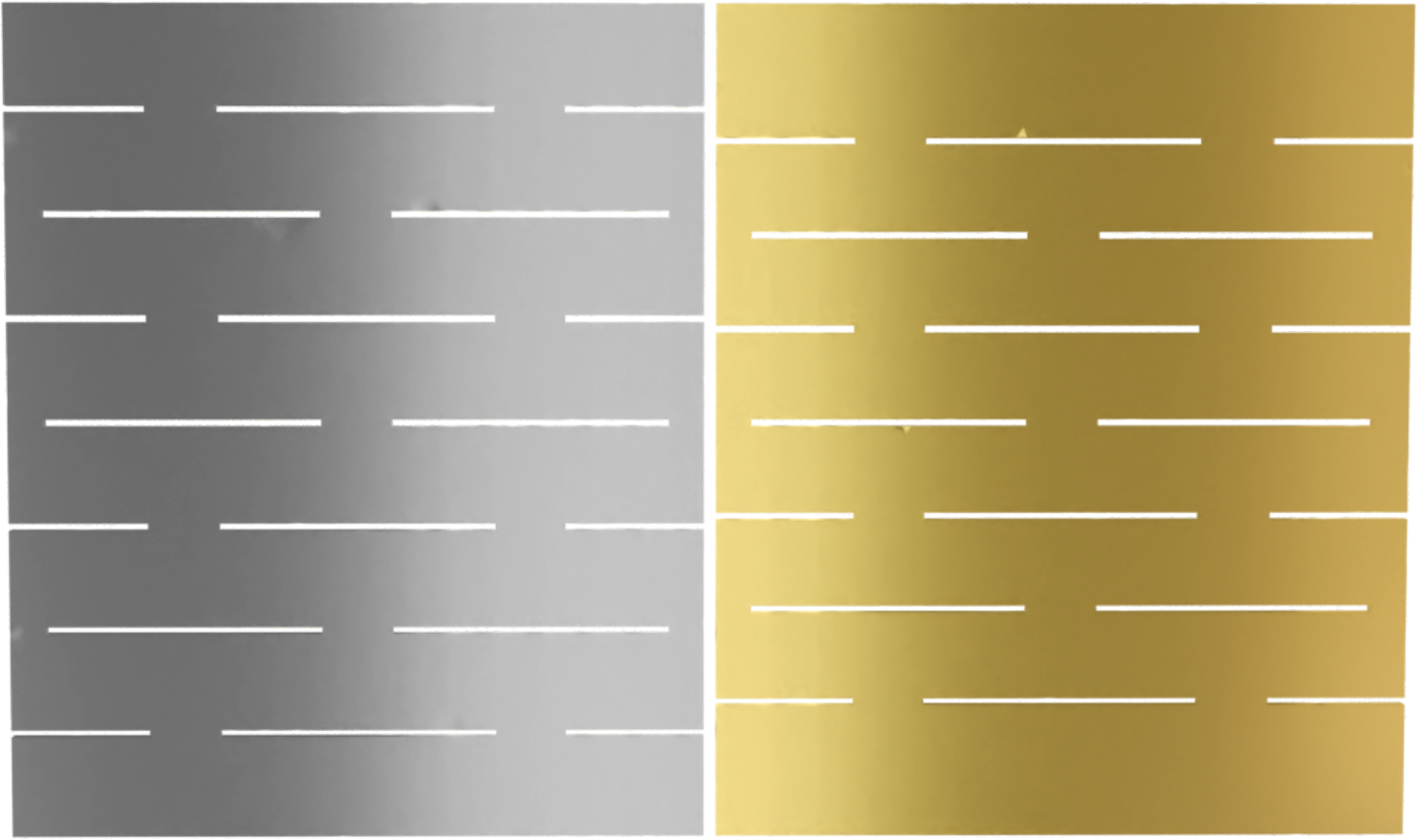}};

    \coordinate (spycenter) at ([xshift=0cm, yshift=-0.6cm] image.center);

    \node[rectangle, draw=none, inner sep=0pt, minimum width=0.73333333333cm, minimum height=0.73333333333cm, anchor=center] (smallspy) at (spycenter) {};

    \node[rectangle, draw=none, inner sep=0pt, minimum width=2.2cm, minimum height=2.2cm, anchor=center] (zoomedspy) at (0,3.8) {};

    \spy [every spy on node/.style={rectangle,draw,black}] 
        on (0,-0.6) 
        in node [] at (0,3.8); 
    \draw[dashed] (smallspy.north west) -- (zoomedspy.south west);
    \draw[dashed] (smallspy.north east) -- (zoomedspy.south east);

    \node[align=center] at ($ (image.south west)!0.25!(image.south east) + (1mm, -1.5mm) $) {Train};
    \node[align=center] at ($ (image.south west)!0.75!(image.south east) + (-1mm, -1.5mm) $) {Test};
    
    \node[align=center, anchor=south] at ($(zoomedspy.north) + (0mm, 0.5mm)$) {Different cuts!};
    
\end{tikzpicture}

\end{minipage}
\hfill
\begin{minipage}[c]{0.75\linewidth} 


    \tikzmarknode{mpStart}{}%
    \vspace{0.5em} \\
    \subfloat[Ground truth (FEM)]{\includegraphics[width=0.19\columnwidth]{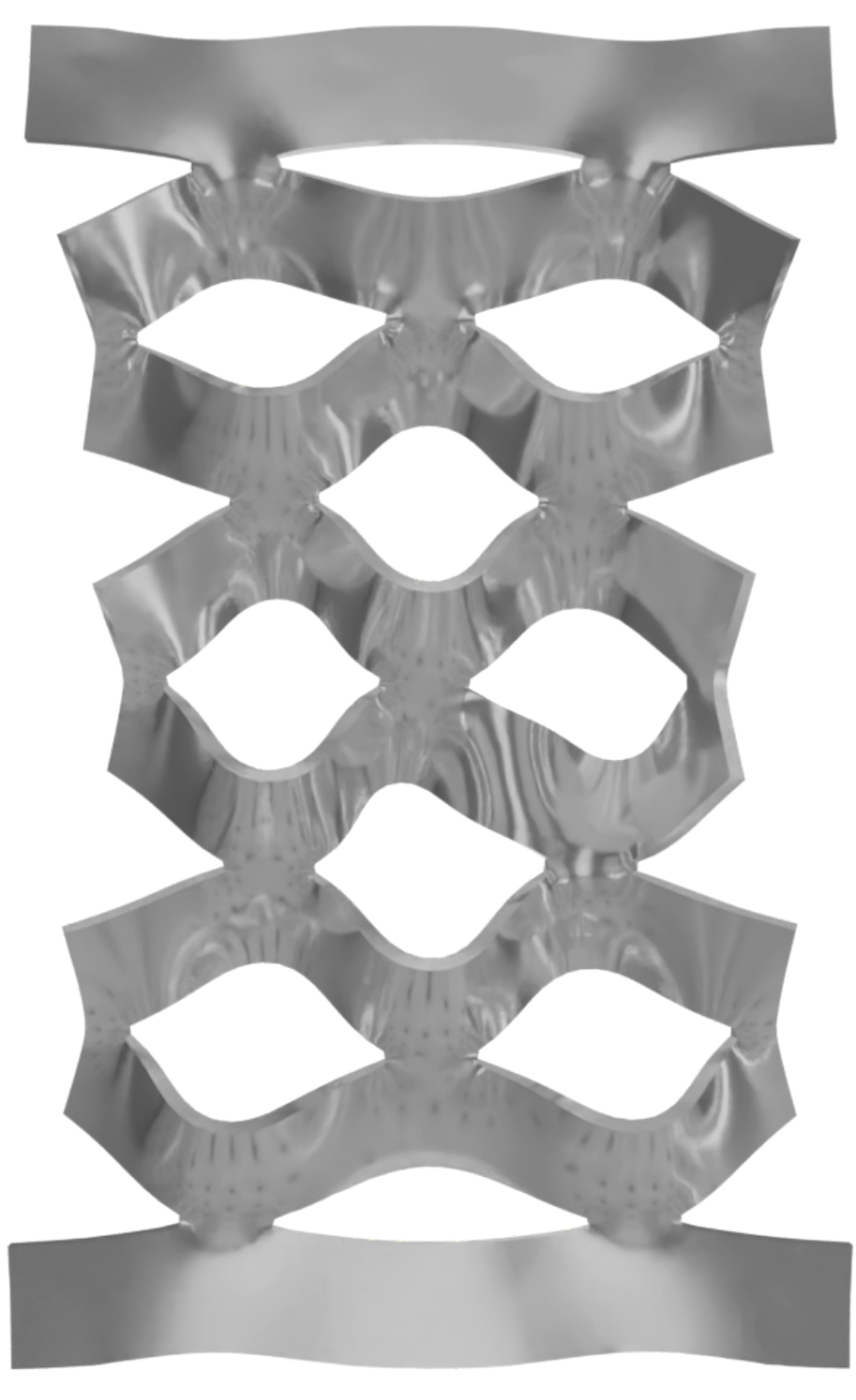}}
    \hfill
    \subfloat[\textbf{Ours \\ \Revision{$\bm{\mathrm{MSE}= 0.02\% }$}}]{\includegraphics[width=0.19\columnwidth]{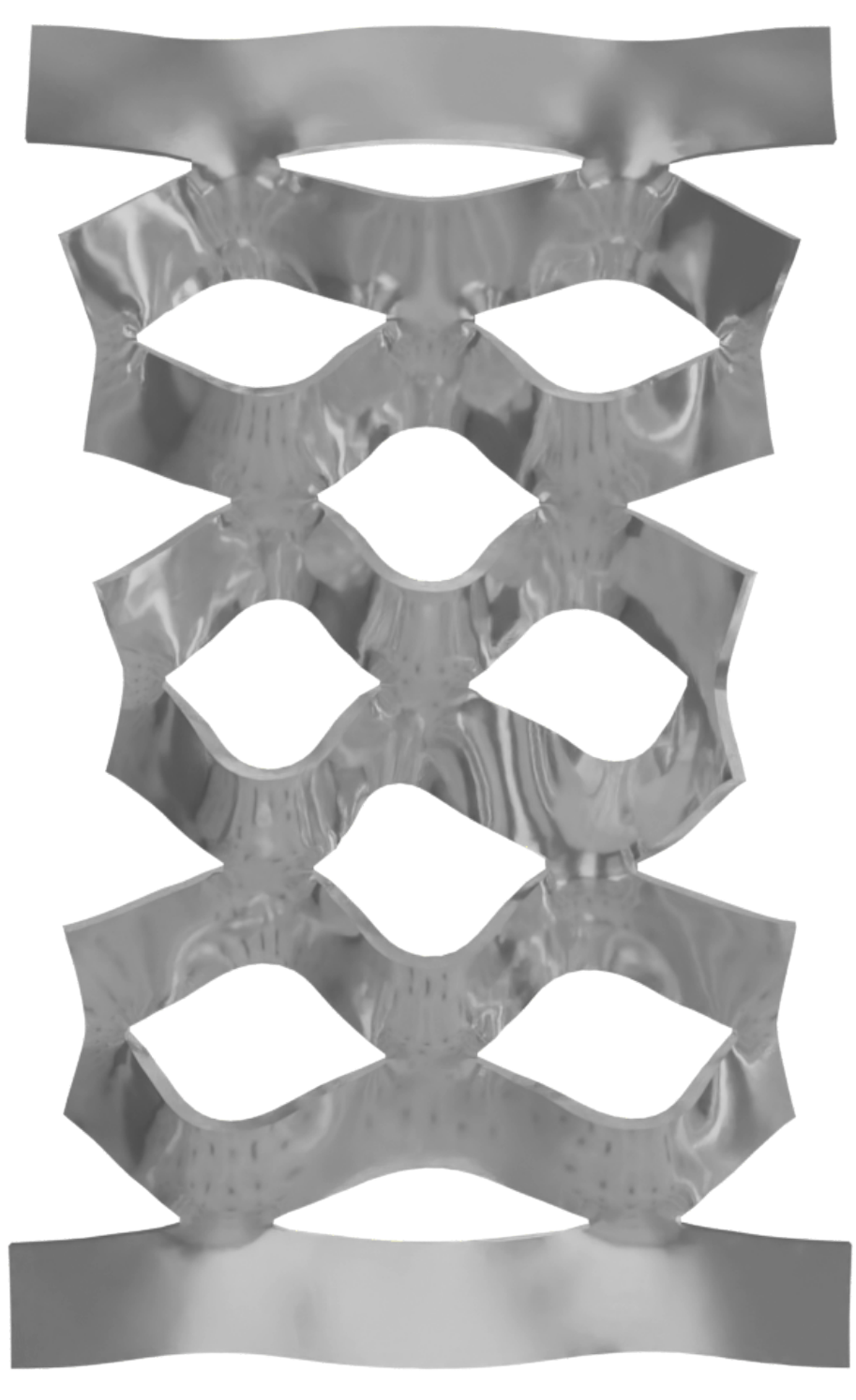}}
    \hfill
    \subfloat[LiCROM \shortcite{chang:2023:licrom} \\
    \Revision{$\mathrm{MSE}= 0.13 \%$}
    ]{\includegraphics[width=0.19\columnwidth]{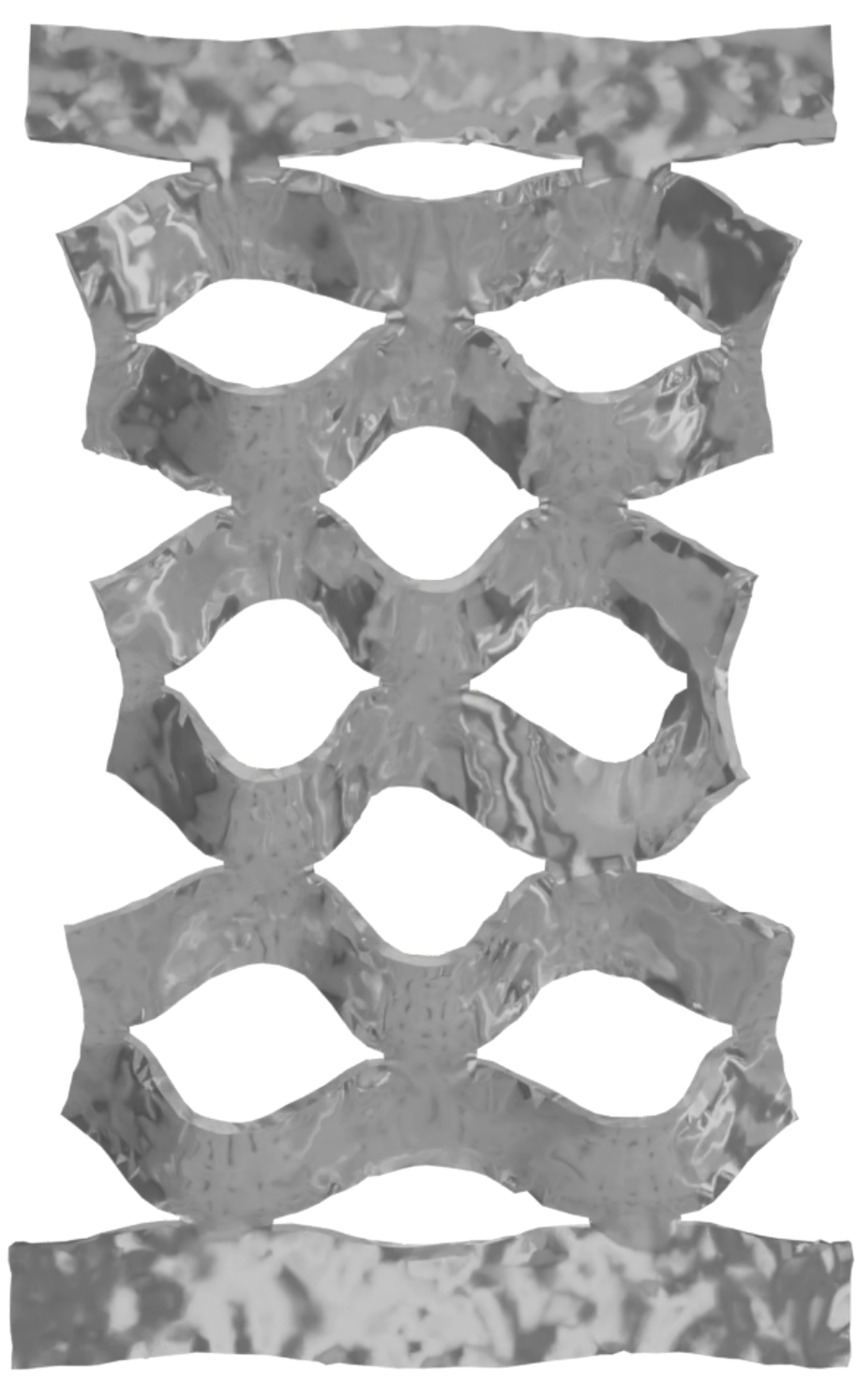}}
    \hfill
    \subfloat[DANN \shortcite{Belhe:2023:DiscontinuityAwareNeuralFields} \\
    \Revision{$\mathrm{MSE}= 0.14 \%$}]{\includegraphics[width=0.19\columnwidth]{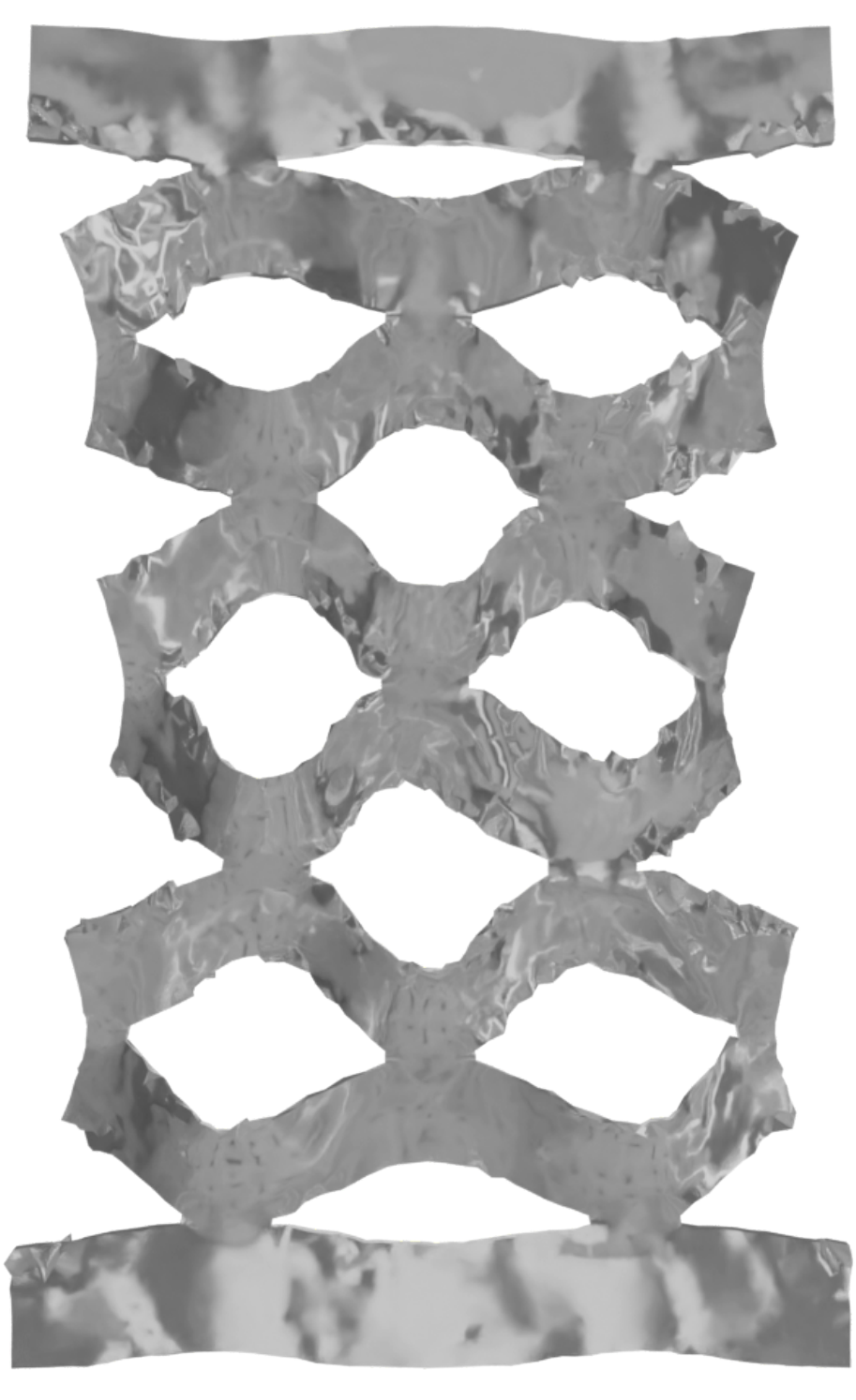}}
    \hfill
    \subfloat[Simplicits \shortcite{Modi:2024:Simplicits} \\
   \Revision{ $\mathrm{MSE}= 3.07 \%$}]{\includegraphics[width=0.19\columnwidth]{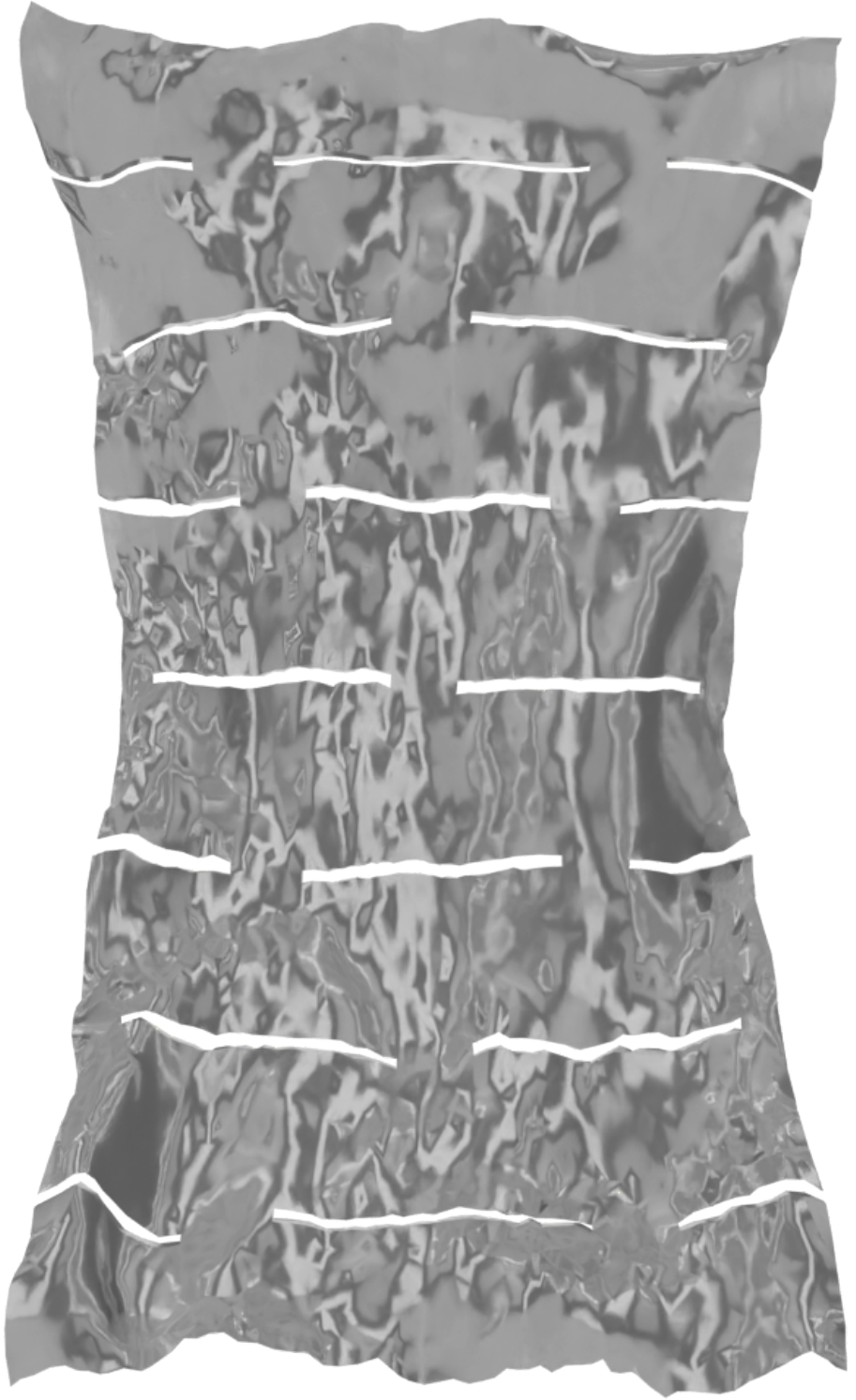}}

    \vspace{1.2 em}
    \subfloat[Ground truth (FEM)]{\includegraphics[width=0.19\columnwidth]{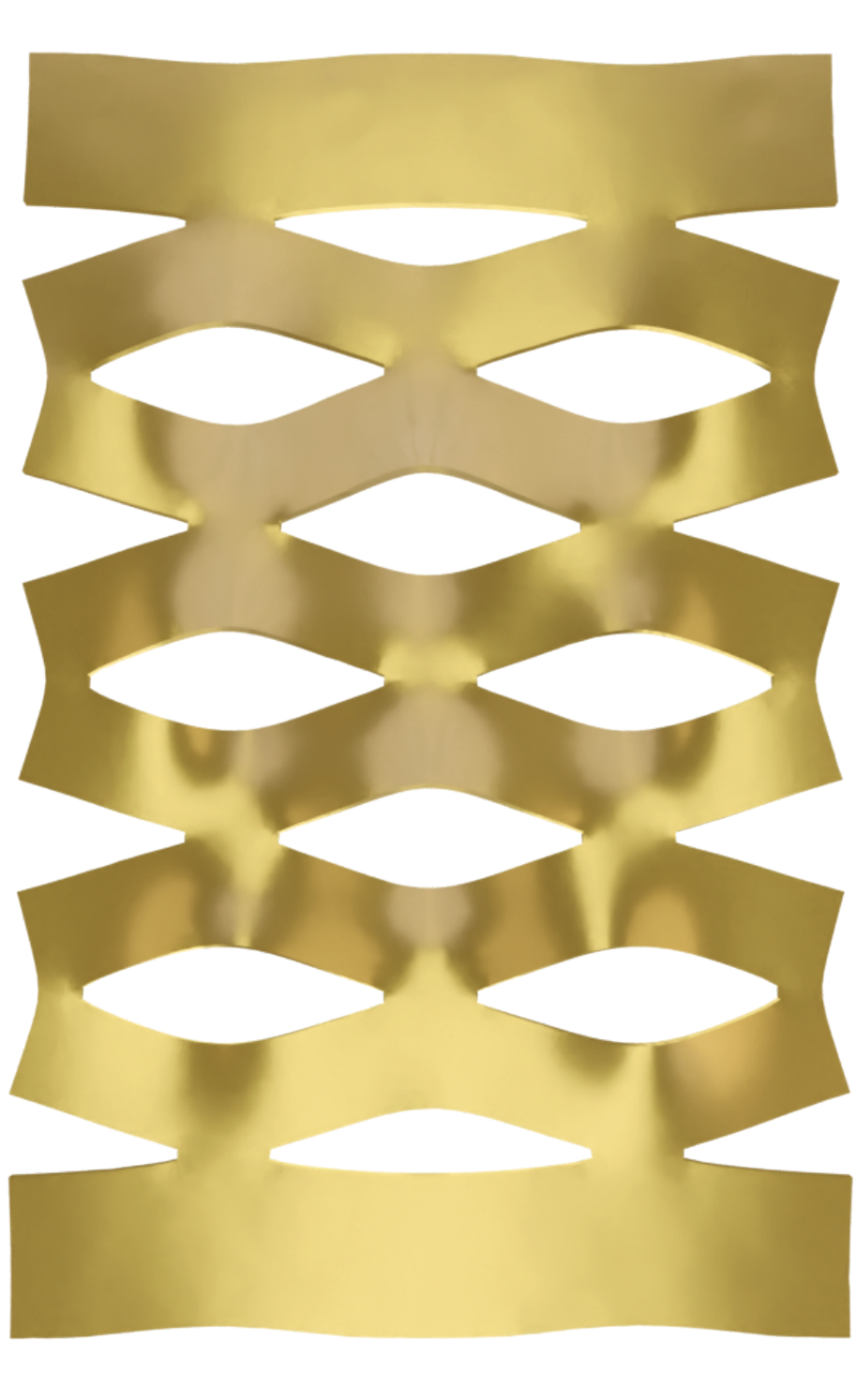}}
    \hfill
    \subfloat[\textbf{Ours \\ \Revision {$\bm{\mathrm{MSE}= 0.20 \%}$}}]{\includegraphics[width=0.19\columnwidth]{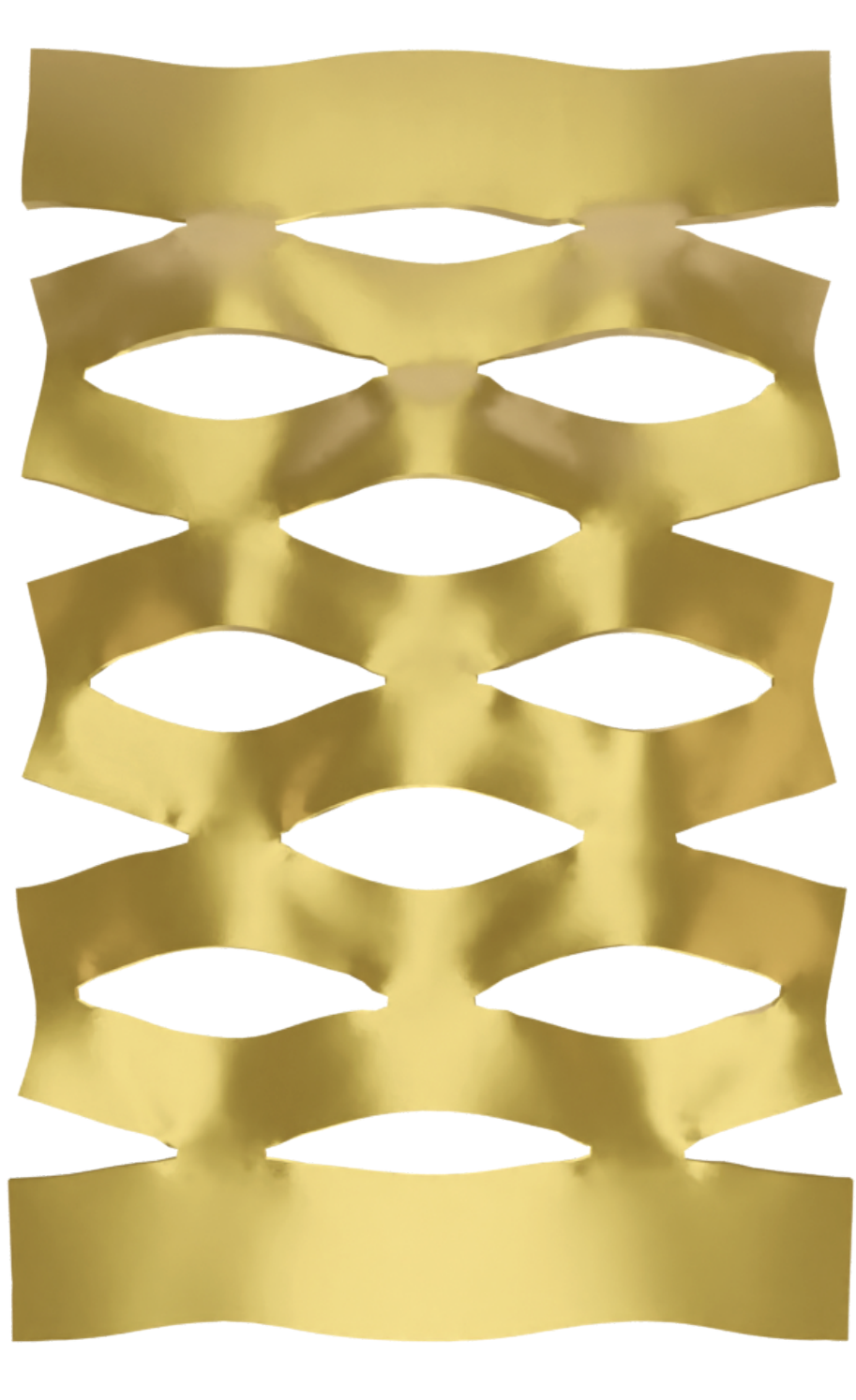}}
    \hfill
    \subfloat[LiCROM \shortcite{chang:2023:licrom} \\
    \Revision{$\mathrm{MSE}= 2.32 \%$}]{\includegraphics[width=0.19\columnwidth]{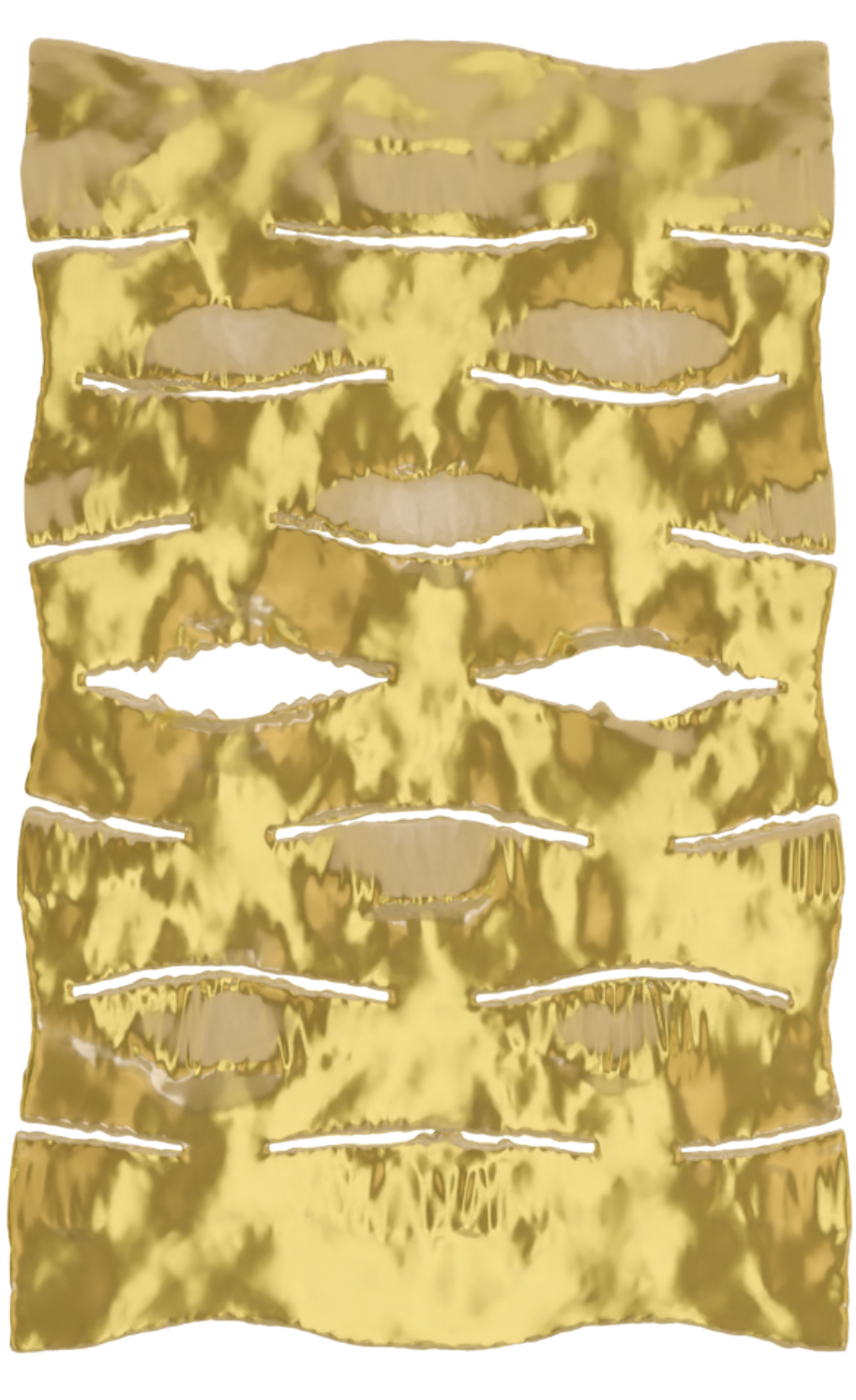}}
    \hfill
    \subfloat[DANN \shortcite{Belhe:2023:DiscontinuityAwareNeuralFields} \\
    \Revision{$\mathrm{MSE}= 2.10\%$}]{\includegraphics[width=0.19\columnwidth]{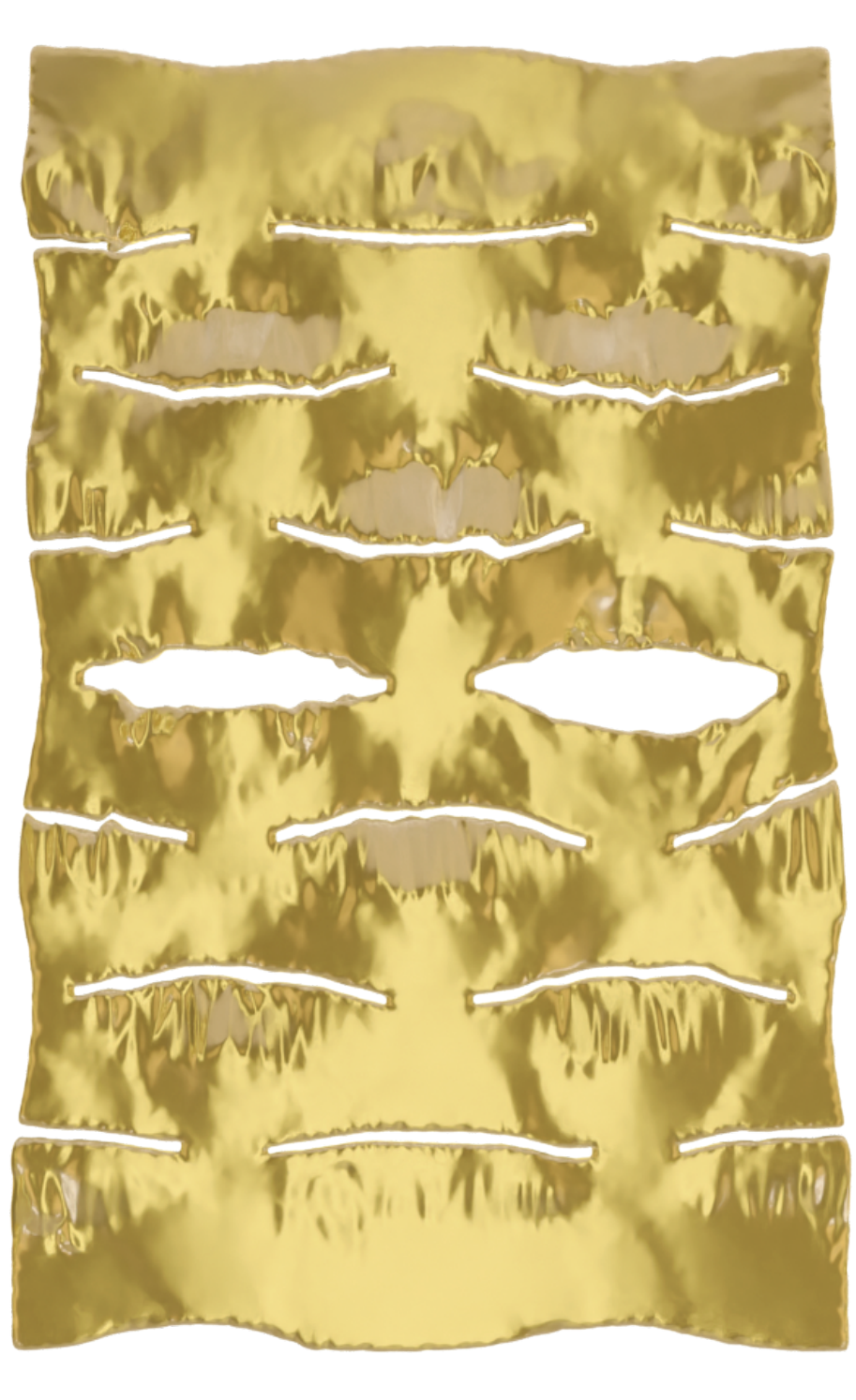}}
    \hfill
    \subfloat[Simplicits \shortcite{Modi:2024:Simplicits} \\
   \Revision{ $\mathrm{MSE}= 2.74\%$}]{\includegraphics[width=0.19\columnwidth]{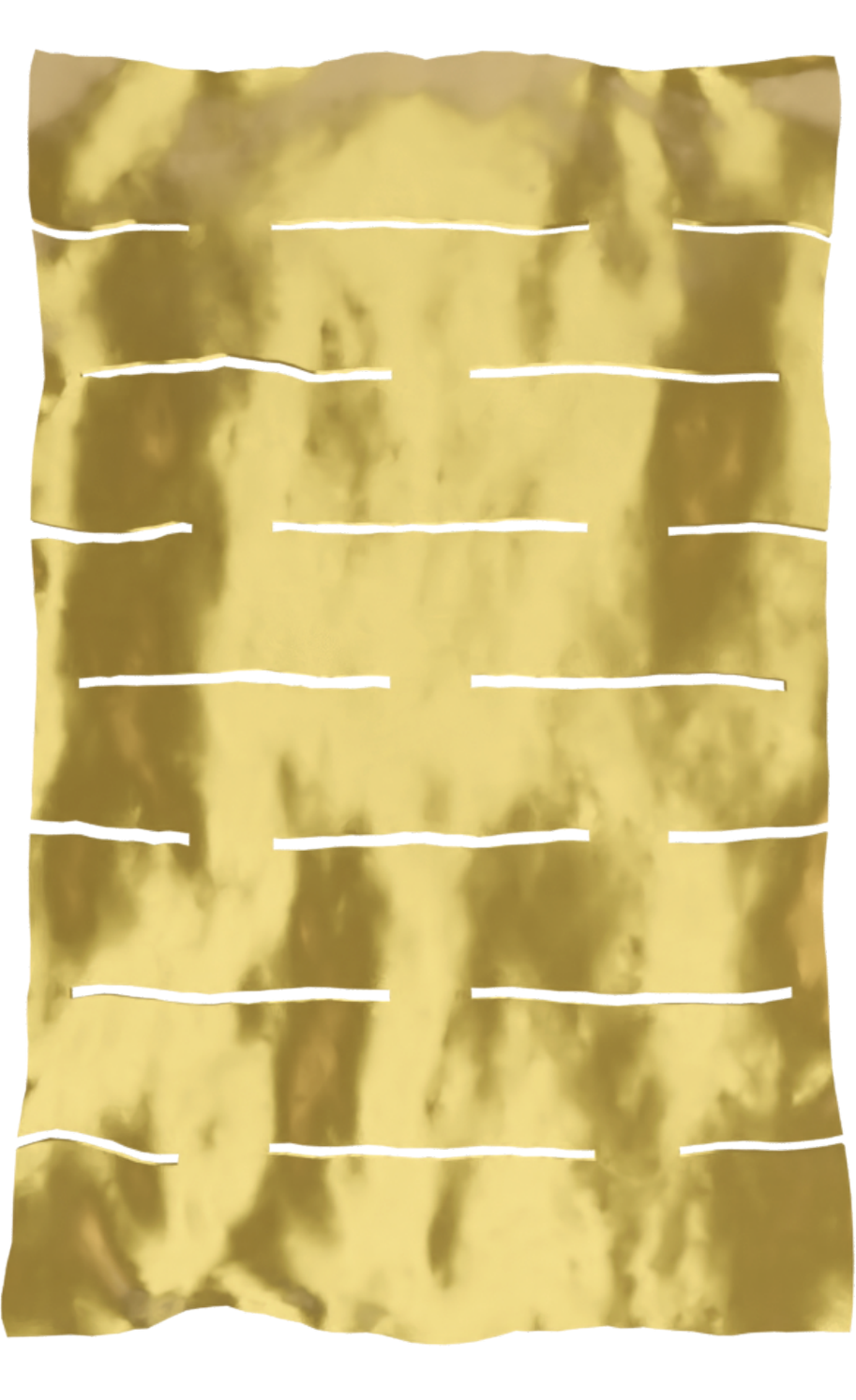}}
  \tikzmarknode{mpEnd}{}%
\end{minipage}
\begin{tikzpicture}[remember picture,overlay]
  \coordinate (mpNW) at ($(mpStart.north west)$);
  \coordinate (mpSE) at ($(mpEnd.south east)$);

    \coordinate (mpNE) at ($(mpNW -| mpSE)$);
    \coordinate (mpSW) at ($(mpSE -| mpNW)$);
    \draw[dashed] ($ (mpNW)!.5!(mpSW) - (0, 1.05em)$) -- ($ (mpNE)!.5!(mpSE) - (0, 1.05em)$);

    \coordinate (mpTopCenter) at ($ (mpNW)!.5!(mpNE) $);
    \coordinate (mpCenter) at ($ (mpNW)!.5!(mpSE) $);

    \node[anchor=north, align=center]
      at ($(mpTopCenter) + (0, 0.45em)$)
      { Shape in training set};

    \node[anchor=north, align=center]
      at ($(mpCenter) + (0, -1.05em)$)
      { Shape not in training set};
\end{tikzpicture}
\caption{
We have compared our \Revision{percentage} reconstruction error with other methods. Our method has the lowest reconstruction error for both shapes in and not in the training set. Moreover, our method is the only one that can generate a visually reasonable result when tested on a cut that is different from the training set.} 
\label{fig:Comparison_results} 
\end{figure*}

As shown in the top row of Figure \ref{fig:Comparison_results}, when evaluated on the training set, the reconstruction errors for all data-driven methods are below \Revision{$0.14 \%$}. Our approach achieves the most accurate reconstruction of deformation details and exhibits the lowest reconstruction error among them. 

\begin{figure*}
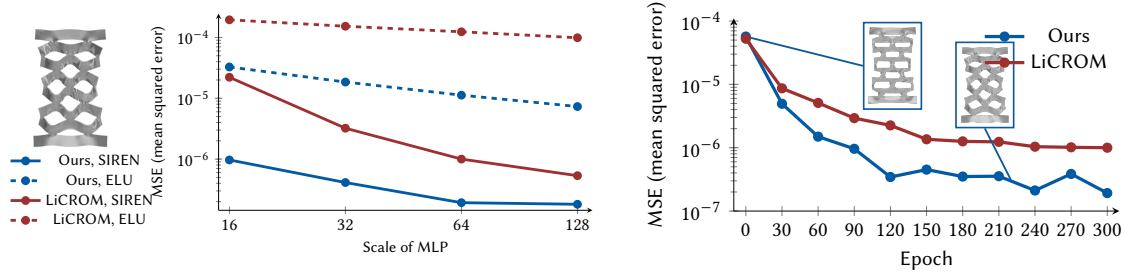

\centering
\begin{minipage}[t]{0.52\linewidth}
    \centering
    \includegraphics[width=\linewidth]{plot_data/tfigure.tikz}
\end{minipage}
\hfill
\raisebox{-0.8em}{
\begin{minipage}[t]{0.43\linewidth}
    \centering
    \includegraphics[width=\linewidth]{plot_data/cfigure.tikz}
\end{minipage}
}
\vspace{-1em}
\caption{We compared the \Revision{percentage} mean squared error \Revision{(MSE)} of our method with LiCROM \cite{chang:2023:licrom} using different MLP scales and activation functions. Our method achieves significantly lower error across all settings, as shown on the left. Additionally, it demonstrates faster convergence during training, as shown on the right. }
\label{fig:combined_mse_conv}
\end{figure*}

To further highlight our method's advantage in modeling discontinuities, we compare the reconstruction error of our method against LiCROM \cite{chang:2023:licrom} across various network scales and activation functions. As shown in Figure \ref{fig:combined_mse_conv}, our method reduces the mean squared error by an order of magnitude under the same settings. Furthermore, our method demonstrates faster convergence compared to LiCROM.

Since Simplicits \Revision{\cite{Modi:2024:Simplicits}} does not require training data, it is expected to have a larger reconstruction error. Simplicits does not consider cuts or discontinuities in its vibration mode analysis, and it also cannot leverage training data that exemplifies the displacements induced by discontinuities. For these reasons, a Simplicits model does not ``know'' that a cut exists. Since the neural field used in Simplicits has built-in continuity, the reduced simulations keep the two sides of the cut glued together. This result reflects the state of the art in geometry-agnostic data-free model reduction until now. By contrast, our approach is explicitly discontinuity-aware, and therefore accurately \Revision{captures} cutting.

When tested on cut shapes not included in training data, our method demonstrates greater generalizability. We have shown the one-shot generalizability of our approach. As shown in the bottom line of Figure \ref{fig:Comparison_results}, when tested on cut shapes with altered cut positions, all previous methods exhibit limitations in reconstructing visually reasonable results. In contrast, our method is flexible enough to adapt to new cut positions by simply updating the $\Gamma$ to the new cut position \emph{without} retraining the neural basis. This adaptation is nontrivial for previous methods because their basis functions are either fully coupled to the neural network weights \cite{chang:2023:licrom} or require remeshing and retraining of the entire domain to handle new cuts \cite{Belhe:2023:DiscontinuityAwareNeuralFields}. 
\Revision{We plot the reconstruction error as a function of the normalized gap difference between the test shapes and those used during training. As shown in Figure~\ref{fig:generalization_curve}, the error increases with the percentage difference between training and test cuts. No significant visual artifacts are observed when the difference is below $15\%$.}

\section{Discussions and Future Work}
In this work, we proposed a novel neural representation technique for capturing precise displacement discontinuities in reduced-order models, with a focus on cuts in thin-walled deformable structures. Our method leverages a generalized winding number field to encode discontinuities, offering significant improvements over traditional mesh-based approaches and other neural field techniques.

\paragraph{Collision} One critical area for improvement is collision handling. While our method successfully represents displacement discontinuities, real-world applications often involve interacting objects where (self-)collision detection and response are critical \cite{zesch2023neural}.

\paragraph{Complex cuts} The cut representation in this paper is limited to piecewise linear curves. Future work could explore extending this approach to incorporate higher-order parametric representations, such as Bézier curves \cite{Spainhour2024_arxiv}.

\begin{figure}
\includegraphics[width=8cm]{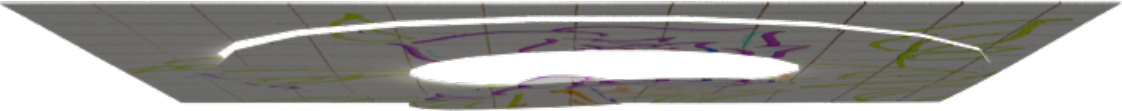}
\caption{When we train our method on a clockwise helical cut and test with a counter-clockwise helical cut, the final deformation is different from the deformation tested on the training set. This is because the winding number distribution is significantly different from the training set.}
\label{fig:failurecase2}
\end{figure}

\paragraph{Out-of-Distribution Challenges} Thanks to the discretization-agnostic approach, our method demonstrates robust generalization capabilities not seen before in subspace physics simulations. However, generalization remains a challenging aspect, particularly for significant, out-of-distribution scenarios. As shown in Figure \ref{fig:failurecase2}, our method's performance decreases when tested on winding number distributions significantly different from the training data. Future research could explore adaptive training strategies or augmentation techniques that ensure the winding number distribution covers a broader range of scenarios \cite{grangier2023adaptive}. Another avenue would be to exploit the volumetric nature of the neural field: it would be interesting to explore whether the field may be trained on multiple alternative cut positions, i.e., supervised by its restriction onto more than one winding graph.


\begin{acks}
 We would like to thank our lab system administrator, John Hancock, and our financial officer, Xuan Dam, for their invaluable administrative support in making this research possible. We acknowledge the support of the Natural Sciences and Engineering Research Council of Canada (NSERC) grant RGPIN-2021-03733.
\end{acks}

\bibliographystyle{ACM-Reference-Format}
\bibliography{main}

\clearpage

\captionsetup[subfloat]{position=bottom,labelformat=empty}

\begin{figure}
\centering
\begin{minipage}[t]{\linewidth}
    \centering
    \includegraphics[width=\linewidth]{plot_data_generalization/cfigure.tikz}
\end{minipage}
\hfill
\caption{\Revision{We compute the percentage mean squared error on cut shapes that deviate progressively from the training set by scaling the cut gap. As the deviation increases, the reconstruction error also rises. When the difference is below 15\%, no significant visual artifacts are observed. Beyond this threshold, while the discontinuity remains accurately captured, deformations begin to appear near the cut edges.}}
\label{fig:generalization_curve}
\end{figure}

\captionsetup[subfloat]{position=bottom,labelformat=empty, skip=0pt, justification=centering}

\clearpage

\end{document}


\title{Supplementary Document for Lifting the Winding Number: \\ Precise Discontinuities in Neural Fields for Physics Simulation}

\author{Yue Chang}
\orcid{0000-0002-2587-827X}
\affiliation{%
  \institution{University of Toronto}
  \country{Canada}}
\email{changyue.chang@mail.utoronto.ca}

\author{Mengfei Liu}
\orcid{0000-0003-4989-6971}
\affiliation{%
  \institution{ University of Toronto}
  \country{Canada}}
\email{mengfei.liu@mail.utoronto.ca}

\author{Zhecheng Wang}
\orcid{0000-0003-4989-6971}
\affiliation{%
  \institution{ University of Toronto}
  \country{Canada}}
\email{zhecheng@cs.toronto.edu}

\author{Peter Yichen Chen}
\orcid{0000-0003-1919-5437}
\affiliation{%
  \institution{MIT CSAIL}
  \country{USA}}
\email{pyc@csail.mit.edu}

\author{Eitan Grinspun}
\orcid{0000-0003-4460-7747}
\affiliation{%
  \institution{ University of Toronto}
  \country{Canada}}
\email{eitan@cs.toronto.edu}

\maketitle

\renewcommand{\theequation}{A.\arabic{equation}} 
\setcounter{equation}{0} 

\section{Integrating the generalized winding number}
Our implementation represents $\Gamma$ as a collection of polylines (allowing for more than one cut). Each polyline $ \mathbf{P} $ consists of $N$ segments. Each segment  $S_i$ is represented by a sequence of points $ \{\mathbf{p}_{i}\}_{i=1}^{N+1} $. 

Since we are going to calculate the integral of the generalized winding number from a partial curve, the segment set that is activated for winding number calculation is changing as the cutting extent $\alpha$ changes. We consider the fractional length of the curve being cut parameterized by $\alpha \in [0, 1]$. Let the segment length of the polyline $l_i$ be defined as $l_i = ||\mathbf{p}_{i+1} - \mathbf{p}_i||$. Then the total length of the polyline is $L = \sum_{i=1}^N l_i$. If $\alpha = 0.25$, this means including segments until the cumulative length reaches 25\% of the total curve length. 

The integration bounds are calculated by walking along the polyline until the total length reaches $\alpha L$. Practically, this means constructing $\Gamma^\alpha$ by sequentially including $\mathbf{p}_i$ from $\mathbf{p}_1$ to $\mathbf{p}_k$ for some $k$, stopping when the total length either meets or would exceed the desired fractional length $\alpha L$  by adding an extra point. If $\sum_{i=1}^k l_i = \alpha L$, then we have obtained our desired polyline that represents $\Gamma_\alpha$, otherwise we would need to partially include the next segment. This is achieved through a linear interpolation. Let $t$ represent the fraction of the last partial segment to be included, calculated as
\begin{align}
    t = \frac{\alpha L - \sum_{i=1}^k l_i}{L_k}.
\end{align}

Therefore, the new endpoint can be calculated as 
\begin{align}
\mathbf{p}^{'}_{k+1}=(1 - t)\mathbf{p}^{k} + t\mathbf{p}^{k+1}.
\end{align}

The new set of $k$ segments with points $ \{\mathbf{p}_{0} \dots \mathbf{p}_{k}, \mathbf{p}'_{k+1}\}_{\alpha} $ can then be used to calculate the generalized winding number $H(\mathbf{x}, \alpha)$ at a query point $\mathbf{x}$.

\begin{align}
\label{eq:discrete_gwn}
    H(\mathbf{x}, \alpha) =  \sum_{i=1}^{k-1} \frac{(\mathbf{x} - \mathbf{p}_{i})^\perp \cdot (\mathbf{p}_{i+1} - \mathbf{p}_{i})}{\|\mathbf{x} - \mathbf{p}_{i}\| \|\mathbf{x} - \mathbf{p}_{i+1}\|} +  \frac{(\mathbf{x} - \mathbf{p}_{k})^\perp \cdot (\mathbf{p}'_{k+1} - \mathbf{p}_{k})}{\|\mathbf{x} - \mathbf{p}_{k}\| \|\mathbf{x} - \mathbf{p}'_{k+1}\|}
\end{align}

\section{Handling Singularities}

The kernel function $K(\mathbf{x}, \mathbf{p}, r)$ is a spatially changing function centered on the end point of the curve $\mathbf{p}=\Gamma(0)$ or $\mathbf{p}=\Gamma(t)$, the radius $r$ of the kernel function is calculated by the distance between the end point and the length of the margin.
\begin{align}
    r &= \|\Gamma(0) - \Gamma(m)\| \quad \text{or} \quad r = \|\Gamma(\alpha) - \Gamma(\alpha-m)\|
\end{align}
We use the following formulation for the kernel function in our implementation:
\begin{equation}
K(\mathbf{x}, \mathbf{p}, r) =
\begin{cases}
\Revision{
   \frac{6\| \mathbf{x} - \mathbf{p} \|^2}{ r^2} \left( 1 - \frac{\| \mathbf{x} - \mathbf{p} \|}{r} \right)}, &  \text{for }  \Revision{0 \leq \| \mathbf{x} - \mathbf{p} \|  \leq \frac{r}{2} } \\[8pt]
 \Revision{1-\frac{1}{4}(2 - \frac{2\| \mathbf{x} - \mathbf{p} \|}{r})^3,} & \text{for } \frac{r}{2} < \| \mathbf{x} - \mathbf{p} \| \leq r, \\[8pt]
1, & \text{for } \| \mathbf{x} - \mathbf{p} \| > r.
\end{cases}
\end{equation}

\begin{wrapfigure}{l}[-3pt]{0.4\linewidth}
\vspace{-10pt}
\includegraphics[width=1.1\linewidth]{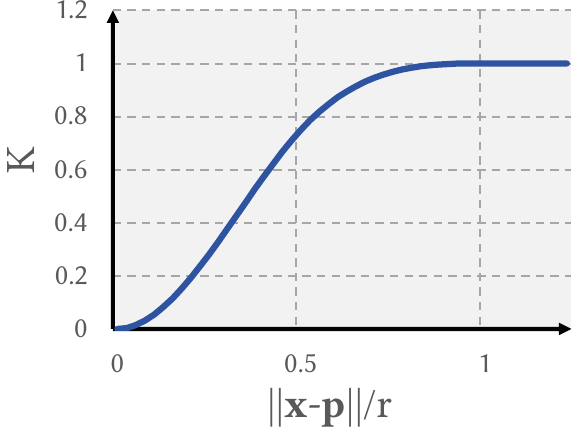}
\vspace{-20pt}
\end{wrapfigure}
This formulation is derived from the cubic spline kernel, modified as one minus the original function to smooth out the singularity at zero distance. \Revision{We visualize the kernel shape on the right. The function is smooth over the entire domain, zero at $\| \mathbf{x} - \mathbf{p} \| = 0$ to cancel the singularity, and constant beyond the margin length $r$.} The final representation of $H$ is therefore: 
\begin{equation}
\label{eq:winding_para_smooth}
\begin{aligned}
    H(\mathbf{x}, \alpha) &= \int_{m}^{\alpha-m} \frac{\Gamma'(t) \cdot (\Gamma(t) - \mathbf{x})^\perp}{|\Gamma(t) - \mathbf{x}|^2} \, dt \\
   & +  K(x, \Gamma(0),  \|\Gamma(0) - \Gamma(m)\|) \int_{0}^{m} \frac{\Gamma'(t) \cdot (\Gamma(t) - \mathbf{x})^\perp}{|\Gamma(t) - \mathbf{x}|^2} \, dt \\
   & +  K(x, \Gamma(\alpha),  \|\Gamma(\alpha) - \Gamma(\alpha-m)\|) \int_{\alpha-m}^{\alpha} \frac{\Gamma'(t) \cdot (\Gamma(t) - \mathbf{x})^\perp}{|\Gamma(t) - \mathbf{x}|^2} \, dt.
\end{aligned}
\end{equation}

\section{Additional Details on Data-free Training}
\label{sec::appendix:training_details}
To construct skin\-ning eigen\-modes \cite{benchekroun2023FastComplemDynamics},
let $\Tilde{w}_\theta : \Omega \times \mathbb{R} \rightarrow \mathbb{R}$ be a continuous volumetric real-valued field, with a corresponding $w: \Omega \rightarrow \mathbb{R} : w = \Tilde{w}_\theta \circ \mathcal{L}$ planar real-valued skinning weight field discontinuous over $\Gamma$. Consider the $k=9$ dimensional reduced space parameterized by the matrix $Z \in \mathbf{R}^{3 \times 3}$, or equivalently by the vector $\bm{z} \in \mathbb{R}^{9} : \bm{z} = \textrm{flatten}(Z)$. Following \citet{benchekroun2023FastComplemDynamics}, the displacement field is 
\begin{align}
    \bm{u}(\bm{x}) = w(\bm{x}) \, Z\, \begin{bmatrix} \bm{x} \\ 1\end{bmatrix} \ .
\end{align}
Observe that $\bm{u}$ is linear in $Z$, therefore, if it were desired, it could be represented in the form $\bm{u} = \bm{z}^T \bm{\Phi}$ described above. Following~\citet{Modi:2024:Simplicits}, we uniformly stochastically sample the 9-dimensional handle $Z$ to train $\Tilde{w}_\theta$ by minimizing the elastic energy 
\begin{align} \label{eq:training-data-free}
    \mathcal{L}_{\text{data-free}} = E_{\text{elas}} = \int_{\Omega}\Psi(\bm{u}(\bm{x})) \, \mathrm{d}\bm{x} 
\end{align}
subject to $\langle w_\theta, 1 \rangle = 1$, i.e., orthogonality to a constant field, a constraint enforced by Gram–Schmidt orthogonalization \cite{chang2024neuralrepresentationshapedependentlaplacian}. We choose a St. Venant-Kirchhoff (StVK) material for the elastic energy density $\Psi$~\cite{barbivc2005real}. The domain integral is estimated using uniform stochastic quadrature. 

This process yields a 9-dimensional basis with a nonuniform weight function $w = \Tilde{w}_\theta \circ \mathcal{L}$. We extend this to an 18-dimensional basis by also incorporating a $3 \times 3$ matrix handle with uniform ($w=1$) weight~\cite{benchekroun2023FastComplemDynamics}.

\bibliographystyle{ACM-Reference-Format}
\bibliography{main}